\documentclass[preprint2]{aastex}

\newcommand{\rsun}{\ \mbox{$\rm R_{\odot}$}}

\newcommand{\msun}{\ \mbox{$\rm M_{\odot}$}}
\newcommand{\mearth}{\ \mbox{$\rm M_{\oplus}$}}

\newcommand{\kms}{\hbox{kms$^{-1}$}}
\newcommand{\ms}{\hbox{ms$^{-1}$}}

\newcommand{\vsini}{\hbox{$v$\ sin\ $i$}}
\newcommand{\vs}{\hbox{$v$\ sin\ $i$}}

\newcommand{\degs}{$\degr$}
\newcommand{\chisqr}{$\chi_r^{2}$}
\newcommand{\chisq}{$\chi^{2}$}

\newcommand{\ha}{H$\alpha$}

\newcommand{\gj}{\hbox{GJ 791.2A}}
\newcommand{\lp}{\hbox{LP 944-20}}

\shorttitle{Starspots on fully convective M dwarfs}
\shortauthors{J. R. Barnes et al.}

\begin{document}

\title{Starspot distributions on fully convective M dwarfs: implications for radial velocity planet searches}

\author{J.R.~Barnes\altaffilmark{1}}
\affil{Department of Physical Sciences, The Open University, Walton Hall, 
       Milton Keynes MK7 6AA, UK}
\email{john.barnes@open.ac.uk}

\author{S.V.~Jeffers\altaffilmark{2}}
\affil{Institut f\"{u}r Astrophysik, Georg-August-Universit\"{a}t, Friedrich-Hund-Platz 1, D-37077 Göttingen, Germany}

\author{H.R.A.~Jones\altaffilmark{3}}
\affil{Centre for Astrophysics Research, University of Hertfordshire, College Lane, Hatfield AL10 9AB, UK}

\author{Ya.V.~Pavlenko\altaffilmark{4,3}}
\affil{Main Astronomical Observatory of the National Academy of Sciences of Ukraine, Golosiiv Woods, Kyiv-127, 03680, Ukraine}

\author{J.S.~Jenkins\altaffilmark{4}}
\affil{Departamento de Astronom\'{i}a, Universidad de Chile, Camino del Observatorio 1515, Las Condes, Santiago. Chile. \\}

\author{C.A.~Haswell\altaffilmark{1} and M.E.~Lohr\altaffilmark{1}}

\begin{abstract}
Since M4.5\,-\,M9 dwarfs exhibit equatorial rotation velocities of order 10 \kms\ on average, radial velocity surveys targeting this stellar population will likely need to find methods to effectively remove starspot jitter. We present the first high resolution Doppler images of the M4.5 dwarf, \gj, and the M9 dwarf, \lp. The time series spectra of both objects reveal numerous line profile distortions over the rotation period of each star which we interpret as starspots. The transient distortions are modelled with spot/photosphere contrast ratios that correspond to model atmosphere temperature differences of \hbox{T$_{\rm phot}$ - T$_{\rm spot}$} \hbox{= 300 K and 200 K}. \gj\ is a {fully convective star with \hbox{\vsini\ = 35.1\ \kms}}.  Although {we find more starspot structure at high latitudes, we reconstruct} spots at a range of phases and latitudes {with a mean spot filling of $\sim3$\%}. \lp\ is one of the brightest known late-M dwarfs, with spectral type M9V and \hbox{\vsini\ = 30.8 \kms}. {Its spectral time series exhibits two dominant transient line distortions that are reconstructed as high latitude spots, while a mean spot filling factor of only 1.5\%\ is found}. The occurrence of low-contrast spots at predominantly high latitudes, which we see in both targets here, is in general likely to be responsible for the low amplitude photometric variability seen in late-M dwarfs. For \gj, the radial velocities induced by the starspot features yield an r.m.s. velocity variability of \hbox{138\ \ms}, which can be reduced by a factor of 1.9 using our reconstructed surface brightness distributions.
\end{abstract}


\section{Introduction}
\label{section:intro}


The Kepler mission has yielded many transiting planet candidates \citep{borucki11kepler,batalha13kepler} while ground based radial velocity (RV) surveys at red or infrared wavelengths \citep{bean10b,barnes12rops} have now demonstrated precision in late-M stars of \hbox{$\sim2.5$ \ms}\ \citep{barnes14rops} indicating the potential to detect rocky planets down to \hbox{$M \sim1.5$\ M$_\oplus$}. For early-M stars, planet occurrence fractions of order 0.5 \citep{bonfils13mdwarfs} and even $>$\ 1 per star \citep{tuomi14mdwarfs} are expected, so understanding any limitations from stellar variability is vital. Meanwhile, M dwarf planet hosting stars have been found to exhibit significant activity with photometric variability of up to $\sim$\ 10\% \citep{knutson11gj436b,crossfield11gj1214b}. Activity will limit searches for low-mass planets around low mass stars in a reasonable number of epochs \citep{barnes11jitter} because by M4.5V stars on average become moderate rotators {  \citep{jenkins09mdwarfs}}, with mean {  \vsini\ \hbox{$\sim5$ \kms}, while rotation is even greater ($\sim 15$\ kms)} for the latest-M stars despite a decline in {\ha}\ activity \citep{mohanty03activity,reiners10activity}.

{Our understanding of magnetic field generation in fully convective stars has only seen advances in relatively recent times. \cite{reiners07magnetic} first observed strong magnetic fields in fully convective stars from Stokes I measurements of FeH lines. A further study \citep{reiners09topology} found that only 14\% of the total magnetic flux detected in Stokes I is seen in Stokes V. Because Stokes I sees the total magnetic field and Stokes V only sees the net magnetic field, this implies that most of the field is arranged on small scales over which opposite polarities cancel each other out. The net field in early M dwarfs, which still possess radiative cores, is nevertheless even lower, at only 6\%. This implies that fully convective stars possess more ordered dipolar field structure in agreement with the Stokes V Zeeman Doppler imaging studies of \citet{donati08mdwarfs}, \citet{morin08mdwarfs} and \citet{morin10mdwarfs}. Donati et al. and Morin et al. found that the topology amongst the earliest M stars (M0V\ -\ M3V) was dominantly toroidal with non-axisymmetric poloidal field structures, while the mid-M (M4V) stars at the fully convective transition boundary exhibited axisymmetric large-scale poloidal fields. The latest stars in the sample (M5\ -\ M8V) were found to possess {either} strongly axisymmetric dipolar fields {or} weak fields with a significant non-axisymmetric component. The stars in these studies did not {however} possess sufficient rotation to enable brightness imaging.}

{Applying Doppler imaging techniques to rapidly rotating M1\ -\ M2 dwarfs, \citet{barnes01mdwarfs} and \citet{barnes04hkaqr} have shown that early-M dwarfs are more uniformly covered with spots than F, G \& K stars. {Images of two M4V dwarfs however reveal fewer spots and low spot filling factors \citep{morin08v374peg,phanbao09}. {  While \cite{crossfield14browndwarf} have published a cloud map of the cool brown dwarf, Luhmen 16B, no images of late M dwarfs have so far been published. In this paper we present Doppler images of the latest fully convective M dwarfs}, using spectral time series observations at red-optical wavelengths.}}

\begin{table*}
\begin{center}
\caption{System parameters and summary of observations for \gj\ and \lp. \label{table1}}
\begin{tabular}{cccc}
\tableline
              & & GJ 791.2A & LP 944-20 \\
\tableline
Spectral Type    &         & M4.5V                                &       M9V                           \\
\vsini\          & [\kms]  & $35.1 \pm {0.4}$                 &  $30.8 \pm {0.5}$                     \\
Axial inclination& [degs]  & ${54 \pm 9}$                     &  $55 \pm {9}$                         \\
Rotation Period  & [days]  & ${0.3088 \pm 0.0058}$            &  $0.162 \pm {0.005}$                \\
$R_*$ (derived)  & [\rsun] & ${0.265 \pm 0.035}$              &  ${0.120 \pm 0.014}$                  \\
$R_*$ (estimated)\tablenotemark{a}  & [\rsun] & $0.279 \pm 0.028$ &  $0.118 \pm 0.012$                  \\
\tableline
N$_{\rm obs}$    &         & 111           & 39             \\        
Exposure times   & [s]     & 180           & 360            \\
Observation span & [MJD]   & 56904.13046   &                \\
2014 Sept 03     &         & - 56904.15523 &                \\
Observation span & [MJD]   & 56906.98665   & 56907.22626    \\
2014 Sept 06     &         & - 56907.21841 & - 56907.40070  \\
                  
\tableline
\end{tabular}
\tablenotetext{a}{Baraffe (1998) assuming 320 Myr (\lp) and 600 Myr (\gj) ages.}
\end{center}
\end{table*}

\nocite{baraffe98}

\subsection{\gj}
\label{section:gj791}
{GJ 791.2A} is an M4.5 dwarf with a previously estimated $v$\ sin\ $i$ = $32.1 \pm 1.7$\ kms$^{-1}$ \citep{delfosse98mdwarfs}. At 8.84\ pc it is a bright, nearby binary system with an astrometrically determined period of $1.4731 \pm 0.0008$ yrs, a maximum separation of $\sim$0.16\arcsec\ {and primary component mass of 0.286\,\msun\  \citep{benedict00gj791}. Our spectra do not show evidence for the secondary \hbox{0.126\,\msun} component, which is estimated to be $\Delta V = 3.23$ fainter than the primary component. \cite{montes01members} found this young disk system does not satisfy Hyades Supercluster membership. 


\subsection{\lp}
\label{section:lp944}

Flaring radio emission has been detected on the M9 dwarf \lp\ \citep{berger01lp944}, which nevertheless exhibits low $L_{{\rm H}_{\alpha}}/L_{\rm bol}$, despite a \vsini\ = 30 \kms\ \citep{mohanty03activity}. \citet{ribas03lp944} determined that \lp\ may be a member of the Castor moving group (\hbox{d = 5\ -20\ pc} from the Sun) with an implied age of \hbox{$320 \pm 80$ Myr}. \citet{dieterich14} report a parallax of \hbox{$155.89 \pm 1.03$\ {mas}} \hbox{(d = $6.41 \pm 0.04$\ pc)}. 

\section{Observations}
\label{observations}

We used the Ultraviolet and Visual Echelle Spectrograph (UVES) at the Very Large Telescope (VLT) to obtain time series spectra of both stars, with a central wavelength of 8300\ \AA, and a spectral range of 6437\ -\ 10253\ \AA. Observations were made with a 0.4\arcsec\ slit, yielding a spectral resolution of $\sim$90,000.
We observed \gj\ on 2014 September 3 \& 6. Since pointing restrictions were in place for the first half of the first night, only 11 spectra of \gj\ were taken, spanning 0.595 hrs (35.7 mins), before telescope closure due to high winds. Both \gj\ and \lp\ were observed on September 6 for \hbox{5.562 hrs} and \hbox{4.187 hrs} respectively. Observations are summarised in Table \ref{table1}.

The spectra were optimally extracted \citep{horne86extopt} using the Starlink package, {\sc echomop}, which removes all but the strongest sky lines. Error information based on photon statistics and readout noise is propagated throughout the extraction process. 

\section{Doppler Imaging}
\label{section:DI}

\subsection{Least squares deconvolution}
\label{section:lsd}

To enable starspot induced line distortions to be {detected} for imaging purposes, we applied {  our implementation of} least squares deconvolution {  \citep{barnes98aper}}. Using this procedure, a high signal-to-noise ratio (SNR) mean line profile is derived for each spectrum with the effects of blending due to rotationally broadened lines removed. The {  deconvolution} algorithm and analysis applied to M dwarfs is detailed in \citet{barnes14rops}. We have used high SNR observations of non-rotating stars of the same or similar spectral type to derive empirical line lists for deconvolution. For \gj, we used GJ 105B (M4.5V) as a slowly rotating standard star, while for \lp, LP 888-18 (M7.5V) was used. 

Obtaining an empirical line list involves removing the blaze, normalising and stitching into a one dimensional spectrum. {Normalisation of spectra is carried out by fitting a spline or polynomial to the spectrum. The minimum number of spline knots or polynomial degree to fit the continuum behaviour is adopted and generally varies with the length of spectrum being fitted. By rejecting outliers at a specified level below each fit and iteratively re-fitting, a good approximation to the continuum level can generally be found. For stars such as M dwarfs, which contain so many lines that only a pseudo-continuum is seen, we found that it was necessary to obtain the maximum value of the spectrum in 10 \AA\ intervals and use these values for continuum fitting.}

Line wavelengths and depths are then identified empirically, with normalised depths $> 0.1$. Lines with strong chromospheric components such as \ha\ and the infrared Na doublet and Ca {\sc ii} triplet are removed. Although LP888-18 has been used successfully to recover precise radial velocities in M7\,-\,M9V stars \citep{barnes14rops}, a closer spectral type match for \lp\ would likely enable line profiles with more optimal SNRs to be obtained. Line lists from model atmospheres do not provide a precise enough match to the observed spectra to enable deconvolution, although in \S \ref{section:models} we use models to obtain line equivalent width ratios and intensity ratios for Doppler imaging. By determining line lists empirically, we include not only atomic lines, but also molecular lines, {which are the dominant source of opacity in cooler atmospheres. For instance, the Vienna Line Database \citep{piskunov95vald,kupka99,ryabchikova11vald} contains $\sim 15800$ lines with normalised depths of 0.5\,-\,1.0 in the \hbox{6500 \AA}\ - \hbox{10250 \AA}\ range for the coolest available temperature of $T =$ 3500\ K. The database currently holds a limited number of molecular species, however for $T =$ 3500\ K $\sim 95$\% of the opacities are due to TiO and MgH (comprising 92\% and 3\% respectively). Of the remaining 5\% atomic metal lines, half have normalised line depths $< 0.165$. By \hbox{3000 K}, spectral energy distributions continue to be dominated by TiO and other diatomic molecular bands, particularly CaH, VO, CrH and FeH \citep{pavlenko14,pavlenko15}. We estimate that 99\% of opacities are due to molecules at $T =$ 3000\ K, which falls to 60\% by $T =$ 2300\ K for atmospheres with 10\,-\,20\% dust. The validity of using temperature sensitive molecular lines for Doppler imaging is investigated in detail in \S \ref{section:models}. }


For \gj, the mean SNR of our input spectra was 57.5. With 7939 lines {  in the \hbox{\hbox{6485\AA\,-\,10253\AA}} region}, the deconvolved profiles possess a mean wavelength of \hbox{$\bar{\lambda} =$ 7781\ \AA}\ and a mean SNR of 3600; an improvement of $\sim63$ in SNR. Similarly, for \lp\ the input and output SNRs were 18.9 and 584 respectively. {  With little flux in the bluest orders, we used the spectral range \hbox{8417\AA\,-\,10253\AA} (\hbox{$\bar{\lambda} =$ 9033\ \AA}) which contains 4150 lines, enabling an effective SNR improvement of $\sim31$ with deconvolution. During deconvolution, telluric lines were used to correct for drifts in UVES following procedures detailed in \citet{barnes14rops}.}

\begin{figure*}
\begin{center}
\includegraphics[scale=0.65,angle=270]{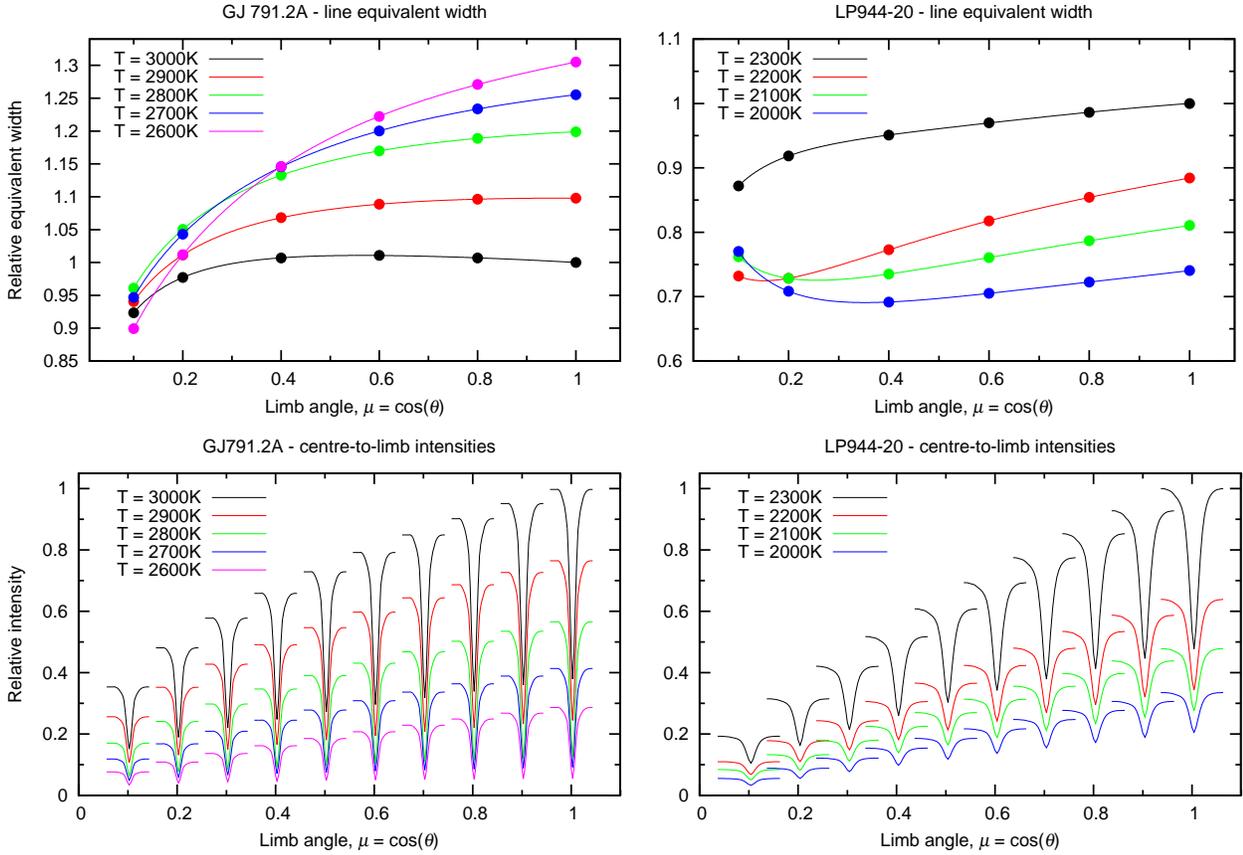} \\
\caption{Top: The deconvolved equivalent width as a function of limb-angle, $\mu$, for BT Settl spectra compute with the WITA code (see \S \ref{section:models} for details). For \gj\ (left), all spectra were deconvolved using a line list with $T$ = 3000 K and for \lp\ (right), using a line list with $T$ = 2300 K. Bottom: The center-to-limb local intensity profiles obtained from deconvolved synthetic spectra. The line profiles are derived from our standard stars, GJ105 (M4.5V) and LP888-18 (M7.5V). The center-to-limb variation of profile equivalent width in the upper panels is adopted. The intensities at each limb-angle, obtained from the synthetic spectra, are the mean intensities determined over the appropriate deconvolution wavelength ranges for our target stars. \label{fig0}}
\end{center}
\end{figure*}

\subsection{Image reconstruction algorithm}
\label{section:dots}

We used the program DoTS \citep{cameron01mapping} to recover Doppler images of our targets. DoTS uses a model with {two brightness levels to derive images} from high resolution, high cadence time series spectra. A spot filling factor, $f_i$, is derived for each pixel, $i$, on the model star by iteratively fitting the time series spectra. 


{Since there are generally more image pixels than data points, and the data possesses finite SNR, often with phase gaps, no unique image can fit the data for a given level of \chisq. A regularising function is therefore imposed on the image. DoTS uses the image entropy

\begin{equation}
S({  f}) = -\sum_{i=1}^{n} w_i \left[ f_i~{\rm ln} \frac{f_i}{m} + (1 - f_i) {\rm ln} \frac{1-f_i}{1-m} \right]
\end{equation}

which has the effect of minimising spurious correlations between image pixels. The function, $S(f)$, combines the entropy, \hbox{$f$\ ln\ $f$}, of both the spot image, ${f}$, and the image \hbox{$(1-f)$\ ln\ $(1-f)$} of the clean photosphere. The default model $m$ is the value that a pixel will assume in the absence of any other constraint imposed by the data. Full details of the image recovery procedure are given in \citet{cameron01mapping}.

To obtain an image solution, the image pixels, $f_i$ are adjusted iteratively to maximise the function

\begin{equation}
Q({  f}) = S({  f}) - \lambda\chi^2({  f})
\end{equation}

where $\lambda$ is the Lagrange multiplier. The value of the Lagrange multiplier is determined such that the image solution lies on the surface with \chisq\ $\simeq$ $M$, where $M$ is the number of measurements in the data set. In practice, systematics, or either underestimated or overestimated data uncertainties mean that that the ideal level of \chisq\ is not achieved. The point at which the rate of change of \chisq\ with each iteration shows a marked decrease indicates a suitable stopping point. Continuing iterations can lead to spot features being recovered that are not justified by the data. A fixed number of iterations is generally adopted for a given data set, which means that the optimal parameters will then provide the best fit (as determined by the minimum achieved \chisq) to the data.}

\subsection{Input models for imaging}
\label{section:models}
 
A study of G, R and I photometric variability in M dwarfs was made by \citet{rockenfeller06mdwarfs} for late M dwarfs. While dust clouds were ruled out for the M9 star 2M1707+64, the light curves were best fit by a dusty atmosphere with cool spots and \hbox{$\Delta T = 100K$}. \cite{berdyugina05starspots} has also shown that $T_{\rm phot} - T_{\rm spot}$ declines with decreasing $T_{\rm phot}$ in giants, leading us to expect contrasts corresponding to temperature differences of only a few hundred K in mid-late M dwarfs. Doppler images of V374 Peg (M4V) were obtained by \citep{morin08v374peg} who employed least squares deconvolution using only atomic line lists and a blackbody temperature difference of \hbox{$\Delta T = T_{\rm phot} - T_{\rm spot}$} \hbox{$= 400$ K}.

Since Doppler imaging requires SNRs of a few hundred to identify starspots in time series spectra and derive reliable images, we combined the signal from as many lines as possible in our observed spectra, including both atomic and molecular lines. Since the behaviour of such a combination of lines with different temperature sensitivities is not clear, we investigated the validity of using least squares deconvolved mean line profiles for Doppler imaging of cool M dwarf atmospheres. 

Synthetic spectra were computed using the BT-Settl model atmospheres \citep{allard12} by the WITA code, first described in \cite{gadun97} using opacity sources listed in \cite{pavlenko07lp944}. Further details can also be found in \cite{pavlenko14} and \cite{pavlenko15}. High resolution disk integrated model spectra were generated in the 2000 K to 3200 K interval in steps of 100 K. In addition, spectra were generated for limb-angles, \hbox{$\mu = cos(\theta)$} \hbox{$= 0.1, 0.2, 0.4, 0.6, 0.8\ \&\ 1.0$}, where $\theta$ is the limb-angle measured from disk center. The spectra were interpolated onto our observation wavelengths to enable us to determine the contrast ratios and line equivalent width behaviour as a function of limb angle over the wavelengths used for deconvolution. For \gj, we adopted  \hbox{$T_{\rm phot}$ = 3000\ K} and for \lp, we adopted \hbox{$T_{\rm phot}$ =} \hbox{2300\ K} in accordance with \citet{reid05} (see also \citealt{kalteneggar09}}).

In the case of \gj, we carried out deconvolution on the synthetic spectra for $T$ = 2600\,-\,3000 K (at 100 K intervals) over the same wavelength range as described in \S \ref{section:lsd}, but using a model-derived line list for the disk integrated spectrum with \hbox{T = 3000 K}. Deconvolution was performed for the spectra at each $\mu$ and the equivalent width of the deconvolved line was measured. We adopted the four-parameter limb-darkening equation used by \citet{claret00ldc4} to fit the line equivalent width variability as a function of limb angle. In addition, we determined the mean intensity of the model spectra at each $\mu$ over the deconvolution wavelengths and again fit the same four-parameter equation to obtain the center-to-limb variation in intensity. We carried out exactly the same procedure for \lp, using a model deconvolution line list for $T$ = 2300 K to deconvolve spectra at all $\mu$ for T = 2000\,-\,2300 K.

Figure \ref{fig0} illustrates the behaviour of the line equivalent width and intensity as a function of limb-angle, $\mu$. The equivalent width at disk center ($\mu = 1.0$) increases monotonically as the temperature decreases from 3000 K to 2600 K. This behaviour is the result of the line opacities becoming stronger at cooler temperatures. By \hbox{2500 K}\ (not shown), the equivalent width begins to decrease again with a relative value of 1.28, (c.f. 1.26 for 2700 K and 1.31 for 2600 K). For the cooler temperature range, with a deconvolution template of 2300 K (Figure \ref{fig0}, top right panel), a monotonic decrease in equivalent width at limb center is instead seen. This behaviour occurs because the line opacities are so numerous and strong that the lines are effectively weakened relative to the pseudo-continuum level. The equivalent width decreases with decreasing limb-angle, except for $T =$\ \hbox{3000 K}, which shows a slight increase out to $\mu$ = 0.5 (Figure \ref{fig0}, top left). Nevertheless, at the cooler temperatures (Figure \ref{fig0}, top right) this trend reverses at the smallest limb angles for $T =$\ 2000 K - 2200 K. While decreases in line equivalent width may result from real changes in line depths, a mismatch between the line list and the spectrum to which deconvolution is applied will lead to non-optimal recovery of the line profile. This effect thus also contributes to trends of the recovered line equivalent width with limb angle and temperature.

The relative center-to-limb variability is illustrated in the lower panels of Figure \ref{fig0}. The intensities at each temperature are relative to the central intensity of the \hbox{3000 K} spectrum (Figure \ref{fig0}, bottom left) and the \hbox{2300 K} spectrum (Figure \ref{fig0}, bottom right). {The line profiles plotted in Figure \ref{fig0} and used for image recovery are however derived from the same slowly rotating standards used in \S \ref{section:lsd} rather than the model spectra, following \cite{morin08v374peg}. Because the standard stars are observed with the same instrumental setup, and because individual line opacities are not a good match, as noted in \S \ref{section:lsd}, we expect that they better represent the local intensity profile {\em shapes} than those derived from the models. The equivalent width from the template star derived line profile is thus adopted for the imagining procedure while the relative equivalent width between the photosphere and spot is obtained from the models (i.e. as indicated in upper panels of Figure \ref{fig0}).}

For a star with a photospheric temperature of $T_{\rm phot}$ = \hbox{3000 K}, the line core is very close to the continuum level at $T_{\rm spot}$ = \hbox{2700 K}, being either above or below depending on the limb angle.  Hence it is likely that spot features will only be distinguishable (in terms of contrast) if the temperature difference between photosphere and spot possess $\Delta T$\ \hbox{$\gtrsim 300$ K}. This corresponds to a contrast at disk center of $I^c_{\rm spot}/I^c_{\rm phot}$ \hbox{= 0.41} for the adopted models with $T_{\rm phot}$ = \hbox{3000 K}. With smaller temperature differences ($\Delta T < 300$ K), we expect the contrast will be too small to enable the spot induced distortions to be adequately fit with a spot filling factors \hbox{$\leq 1.0$}. Similarly for a star with \hbox{$T_{\rm phot}$ = 2300 K}, a temperature difference of \hbox{$\Delta T$\ $\sim 200$ K} ($I^c_{\rm spot}/I^c_{\rm phot}$ \hbox{= 0.48}) is likely required to enable spot filling factors \hbox{of 1.0}.  In section \S \ref{section:results}, we present results for image recovery using spot/temperature contrasts corresponding to $\Delta T$ \hbox{$= 300$ K} and $\Delta T$\ \hbox{$= 200$ K} for \gj\ and \lp\ respectively and additionally investigate the effect of modifying the adopted contrasts.

\begin{figure*}
\begin{center}
\includegraphics[scale=0.65,angle=270]{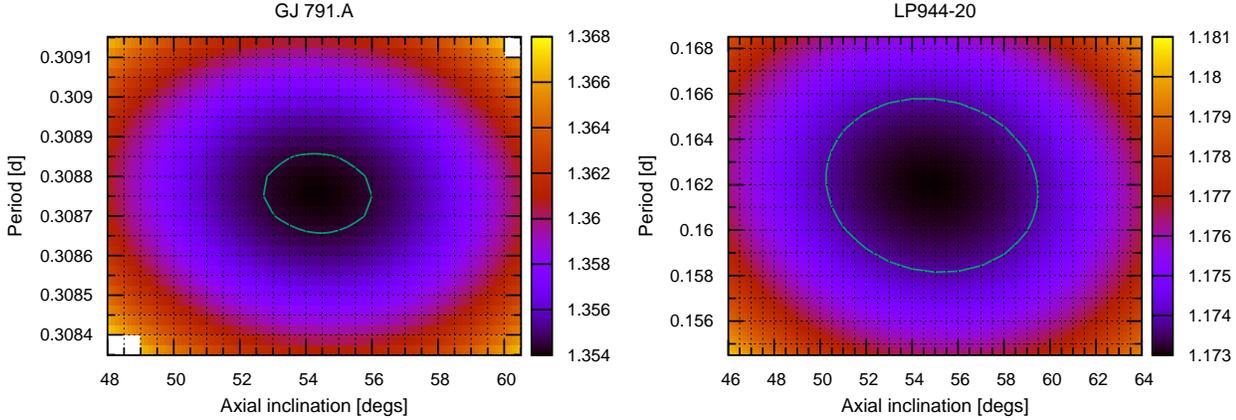} \\
\caption{{Reduced \chisq\ (\chisqr)} plots of rotation period, $P$, vs axial inclination, $i$ with 67.3\% confidence level. \chisqr$_{\rm min}$ = 1.354 (\gj) and \chisqr$_{\rm min}$ = 1.173 (\lp). The derived parameters and uncertainties are listed in \hbox{Table 1}.\label{fig1}}
\end{center}
\end{figure*}

\begin{figure*}
\begin{center}
\includegraphics[height=170mm,angle=270]{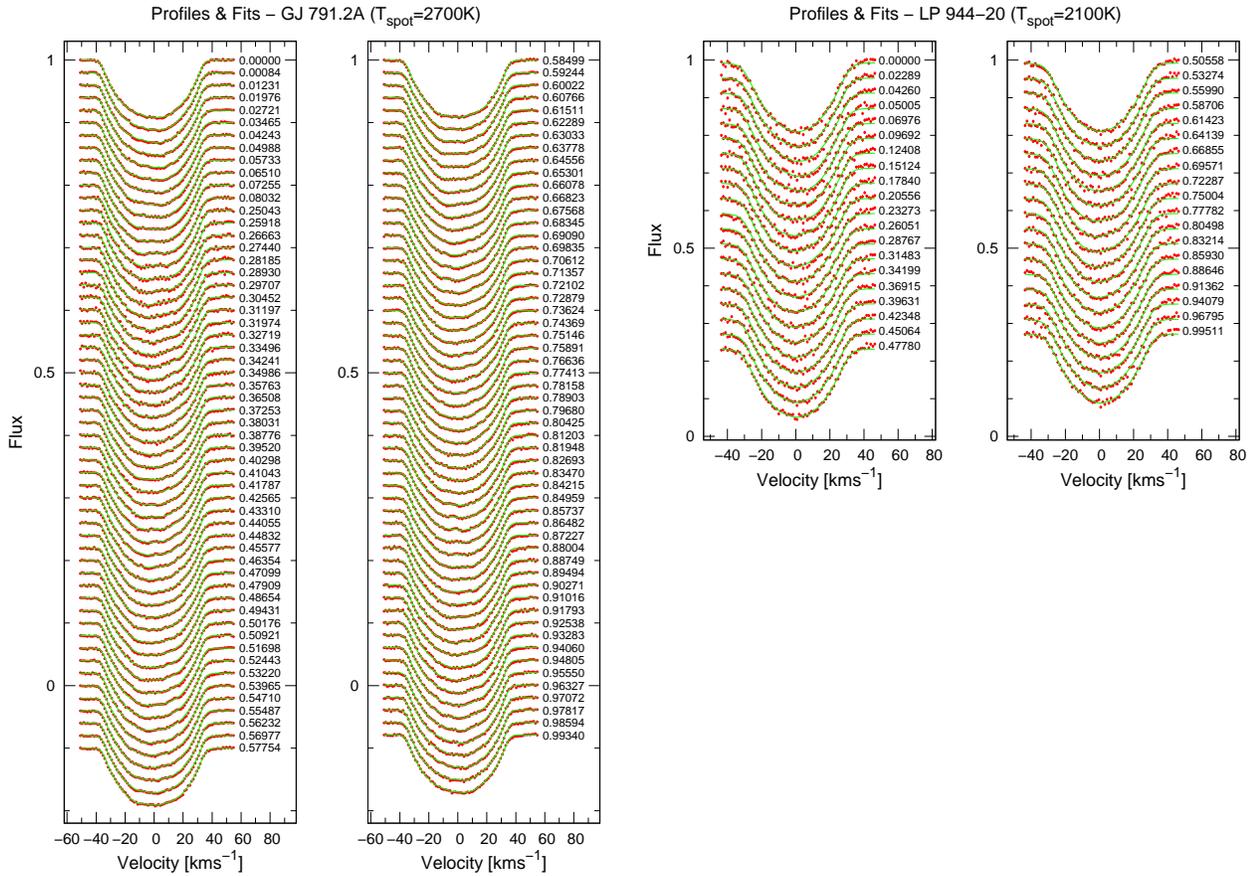} \\
\end{center}
\caption{Deconvolved profiles for \gj\ (left) and \lp\ (right). The data are plotted as points along with the error bars. The maximum entropy regularised profiles corresponding to the images in Figures \ref{recongj791} \& \ref{reconlp944} are shown as line plots. Observations are phased with epochs HJD0 = 2456904.63601 (\gj) and HJD0 = 2456907.73034 (\lp), which are the mid-exposures of the first observation in each time series. \label{decon-timeseries-both}}
\end{figure*}

\begin{figure*}
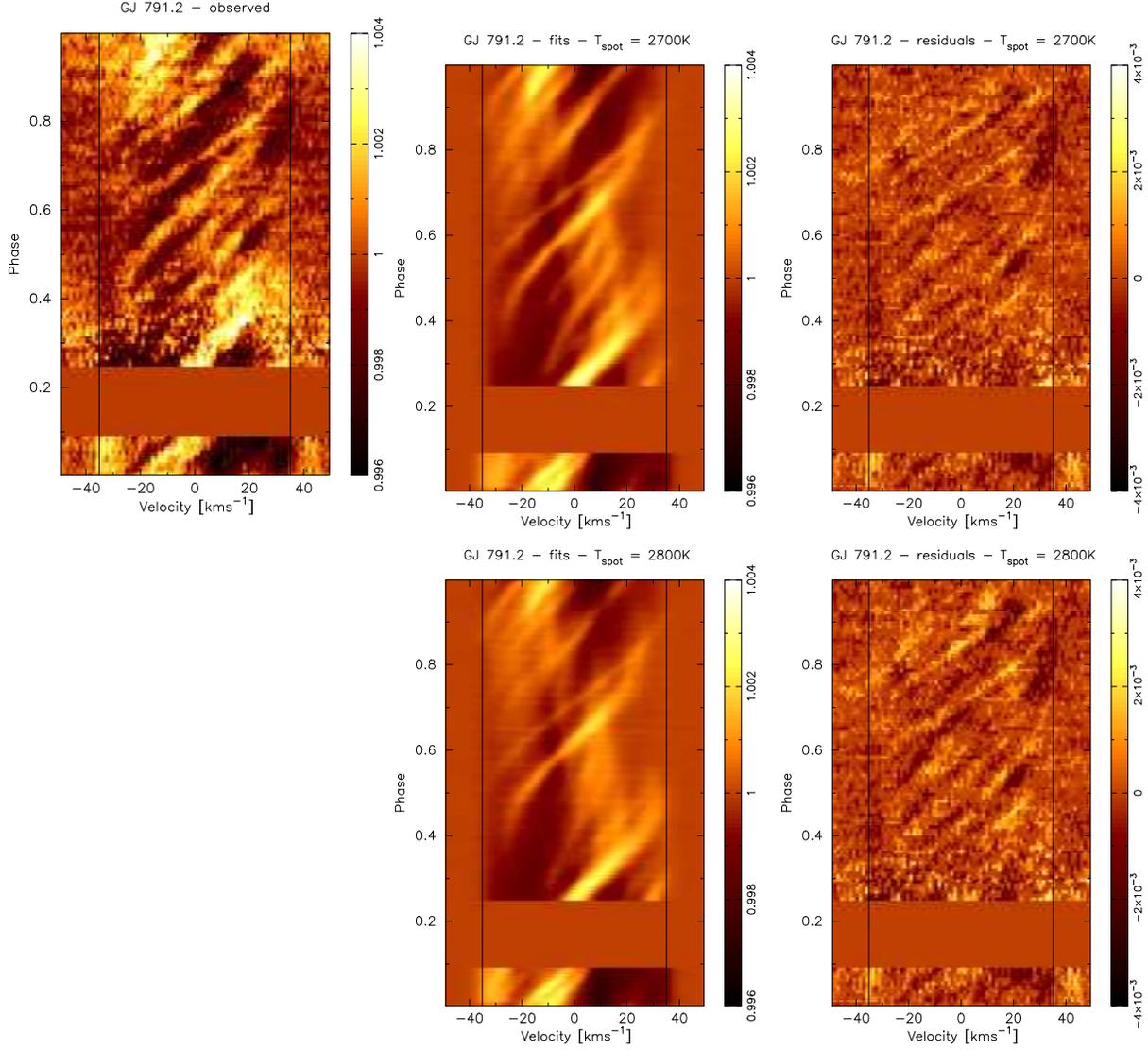

\begin{center}
\begin{tabular}{ccc}
\includegraphics[width=52mm,angle=0,bbllx=0,bblly=0,bburx=524,bbury=765]{timeseries_observed_gj791-jun03.ps} \hspace{-3mm} &
\includegraphics[width=52mm,angle=0,bbllx=0,bblly=20,bburx=544,bbury=765]{timeseries_fits_gj791-28its-2700K-jul29.ps} \hspace{-3mm} &
\includegraphics[width=52mm,angle=0,bbllx=0,bblly=20,bburx=544,bbury=765]{timeseries_residuals_gj791-28its-2700K-jun03.ps} \\ 
                                                                                                                                    &
\includegraphics[width=52mm,angle=0,bbllx=0,bblly=20,bburx=544,bbury=765]{timeseries_fits_gj791-28its-2800K-jul29.ps} \hspace{-3mm} &
\includegraphics[width=52mm,angle=0,bbllx=0,bblly=20,bburx=544,bbury=765]{timeseries_residuals_gj791-28its-2800K-jun03.ps} \\ 
\end{tabular}
\end{center}
\caption{Mean profile divided deconvolved line profile time series (left), fits (middle) and fit residuals (right) for \gj\ using a spot temperature of T$_{\rm spot} = 2700K$ (upper panels) and T$_{\rm spot} = 2800K$ (lower panels). The starspot features appear as white trails while black trails are regions that are more negative relative to the mean profile. The spectra are centered on the observed rest frame and phased according to the adopted 0.3088\ d period of \gj. The vertical lines indicate a \vsini\ profile width of of \hbox{35.1 \kms}. \label{timeseriesgj791}}
\end{figure*}

\begin{figure*}
\begin{center}
\begin{tabular}{ccc}
\includegraphics[width=52mm,angle=0,bbllx=0,bblly=20,bburx=544,bbury=765]{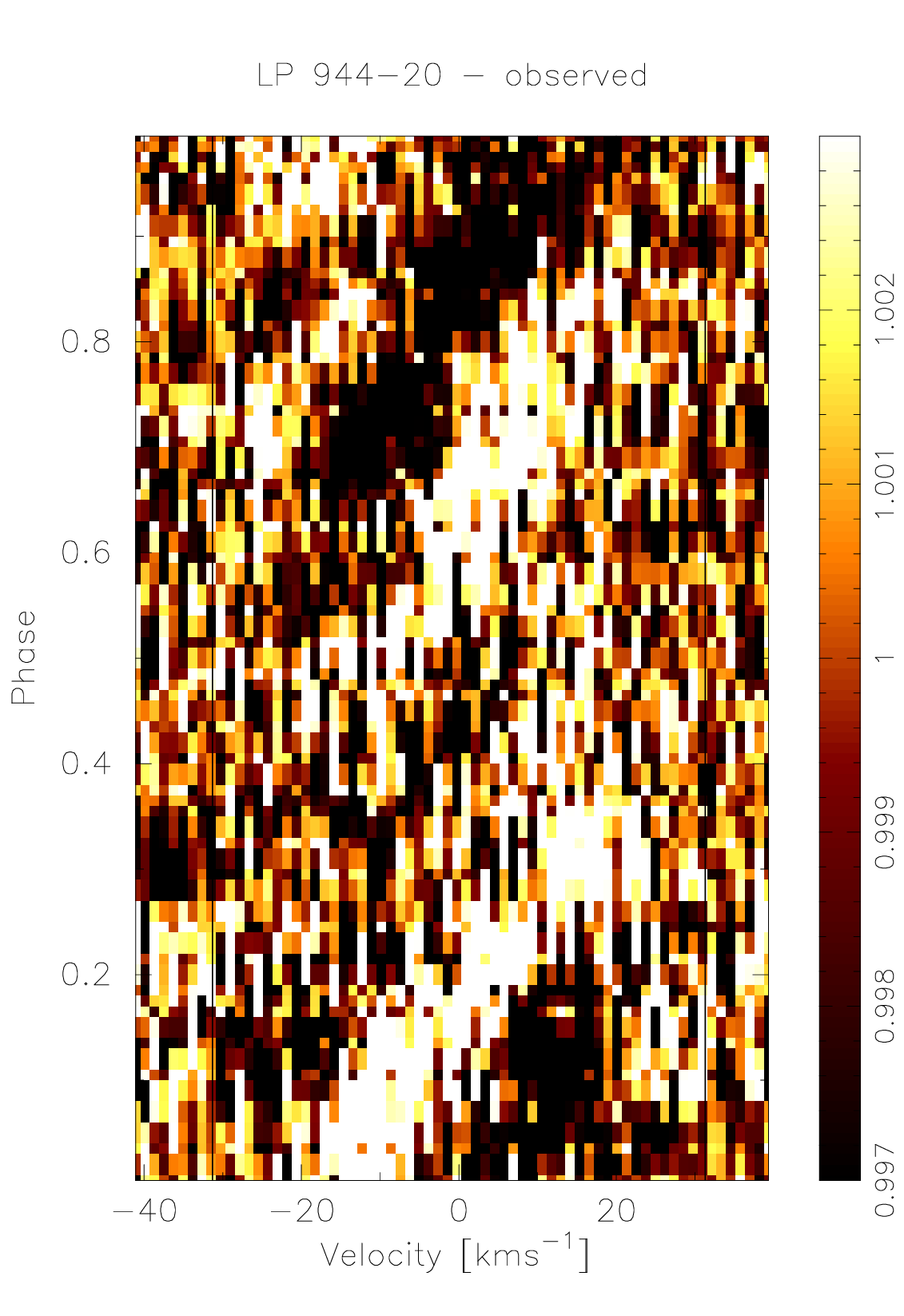} \hspace{-3mm} &
\includegraphics[width=52mm,angle=0,bbllx=0,bblly=20,bburx=544,bbury=765]{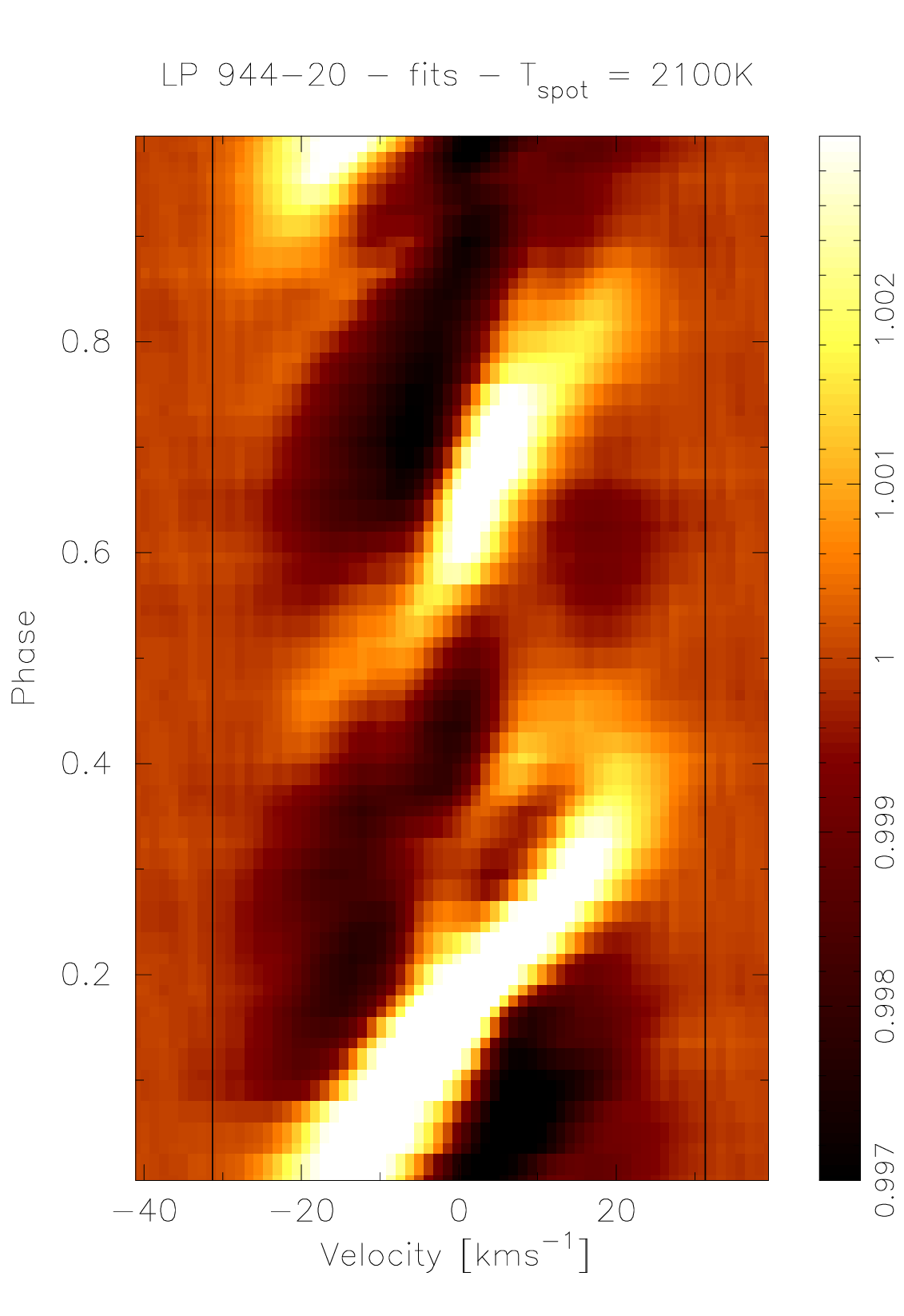} \hspace{-3mm} &
\includegraphics[width=52mm,angle=0,bbllx=0,bblly=20,bburx=544,bbury=765]{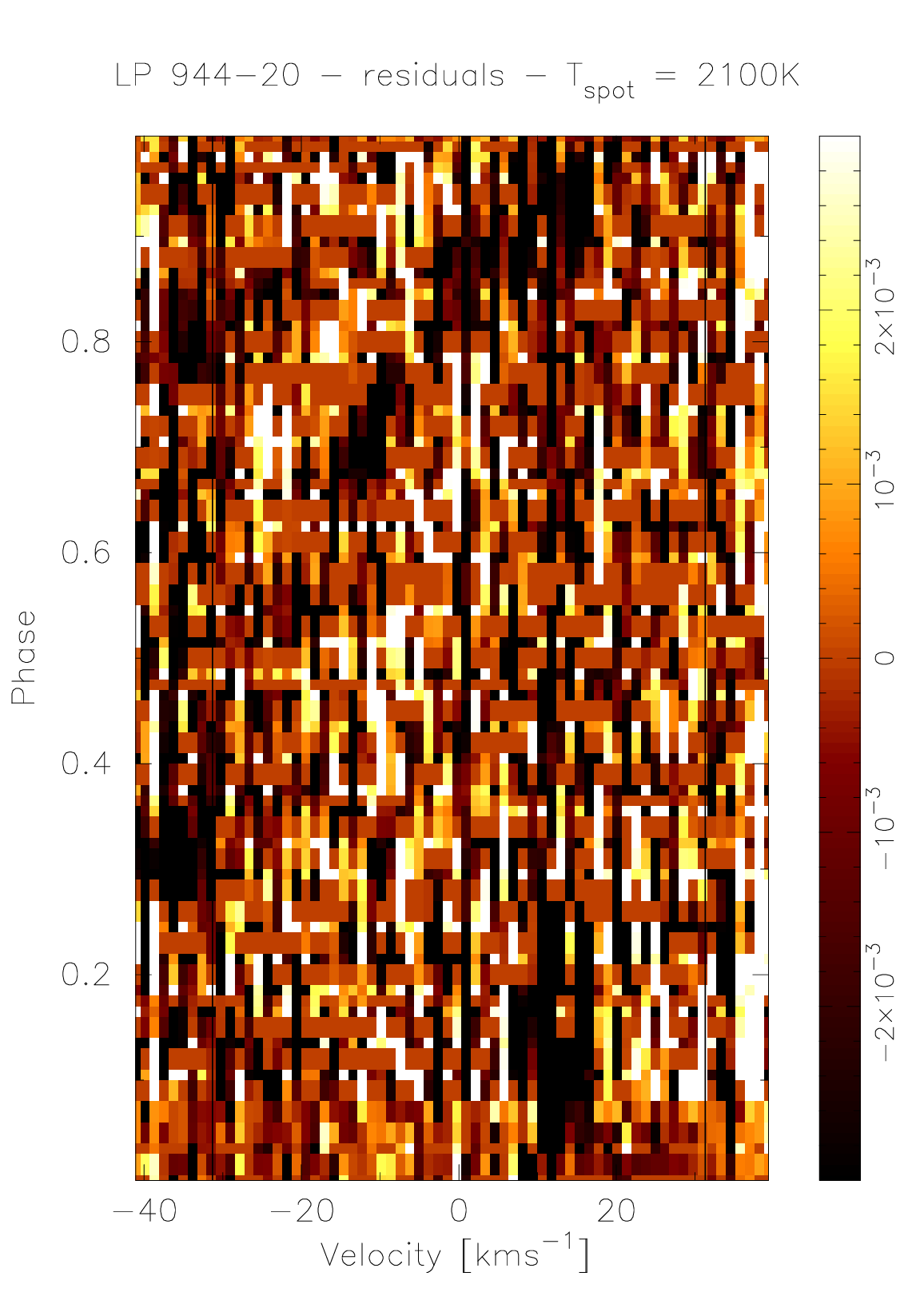} \\ 
                                                                                                                                    &
\includegraphics[width=52mm,angle=0,bbllx=0,bblly=20,bburx=544,bbury=765]{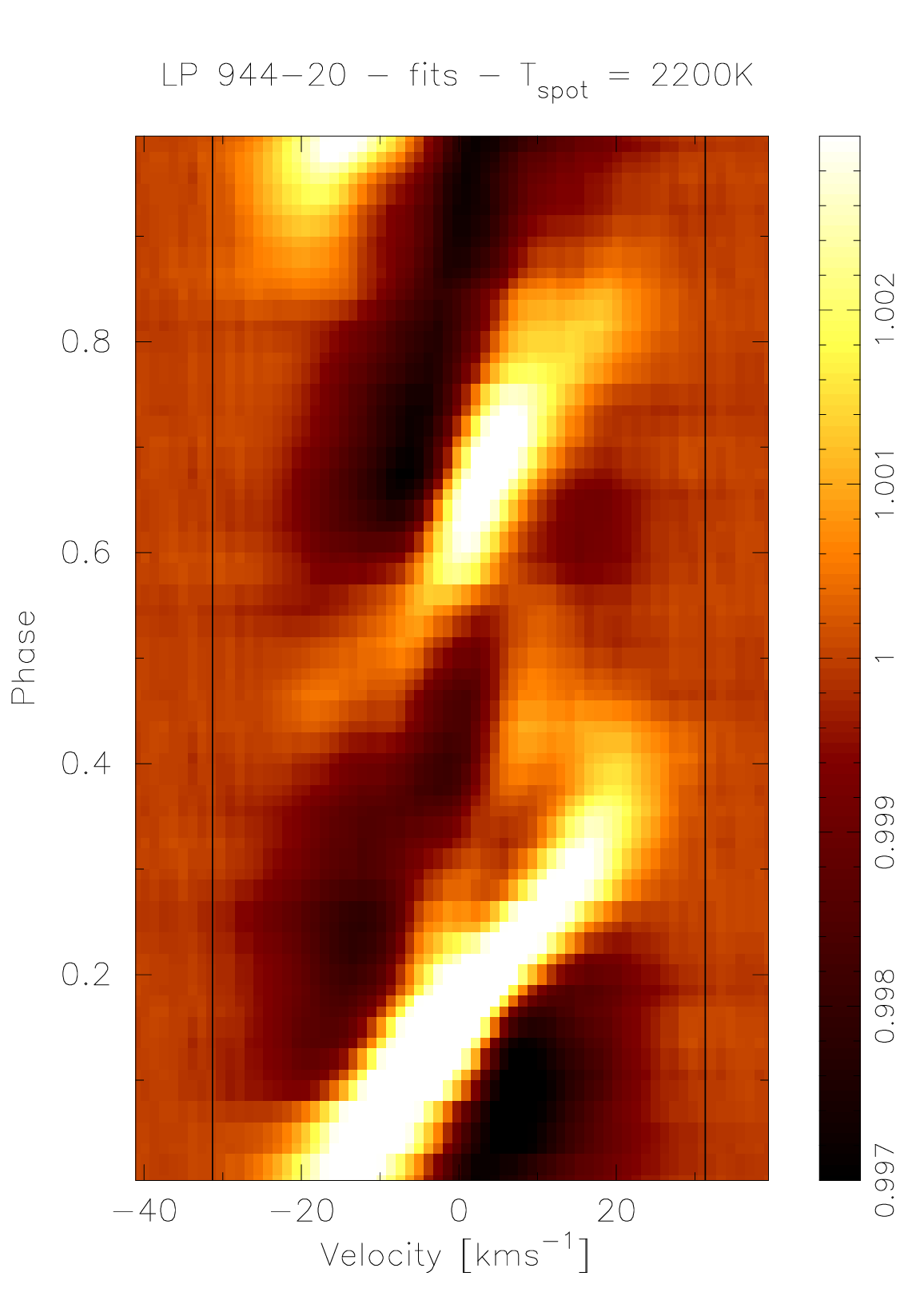} \hspace{-3mm} &
\includegraphics[width=52mm,angle=0,bbllx=0,bblly=20,bburx=544,bbury=765]{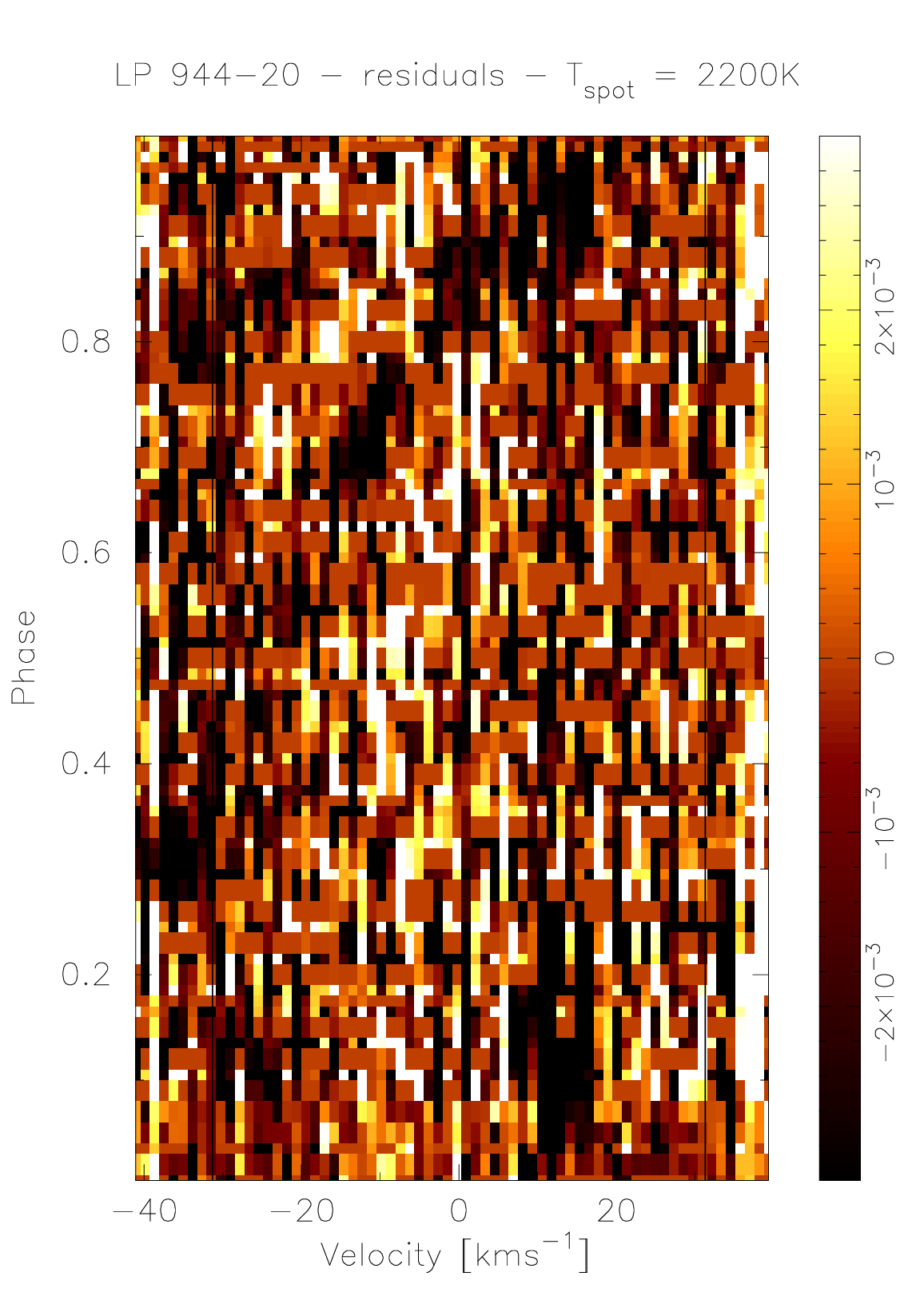} \\ 
\end{tabular}
\end{center}
\caption{As for Figure \ref{timeseriesgj791}, but for \lp. \label{timeserieslp944}}
\end{figure*}

\subsection{Rotation periods and system parameters}
\label{section:system}

%

{  The SuperWASP archive \citep{butters10wasp}\footnote{http://www.superwasp.org} contains 10166 observations of GJ 791.2A, with mean photometric errors of 0.24, spanning $\sim$6.5 years. SuperWASP is however not sensitive to objects, such as \lp, with $V \gtrsim15.0$. We searched for periodicity in the data using a phase dispersion minimisation algorithm \citep{lohr14}, obtaining \hbox{$P =  0.257025(1)$\ d}. 
The amplitude of variability, \hbox{$\Delta V = 0.02$} (8\% of the photometric errors), is comparable with the amplitudes found by \citet{rockenfeller06mdwarfs} for other variable M dwarfs. However, folding subsets of data spanning only a few tens of days does not reveal obvious periodicity, suggesting higher cadence data with better photometric precision is needed.

Adopting published parameters can lead to systematic and characteristic biases in the reconstructed images \citep{rice89,unruh95profile}.  For example, the rotation velocity, \vsini, and line profile depth combination are correlated, and the optimal \chisq\ for a given input model must be determined \citep{barnes00pztel}. Over-estimation of profile equivalent width typically leads to images with completely filled polar regions as the fitting procedure attempts to match the observed profiles by reducing the depth of the line center. This can be seen for example in the images of G dwarfs presented by \cite{barnes98aper}, as compared with the same images in \cite{barnes01aper}, where a lower equivalent width and \vsini\ were adopted following a revised parameter optimisation procedure, which we use here. The main system parameters, recovered by minimising \chisq\ in the the multi-dimensional parameter space are listed in Table \ref{table1}. 

Uncertainties in parameters can be obtained from the likelihood function based on the minimum \chisq\ values obtained and the number of data points contributing to the image solution, as outlined in \S \ref{section:dots}. {We have also considered the effect of systematic errors that could arise from adopting the wrong local intensity profiles and the $\Delta T$ values and limb darkening parameters investigated in \S 4. For both \gj\ and \lp, using the model-derived local intensity profiles leads to \vsini\ values that are 0.2 \kms\ lower than with the adopted standard star local intensity profiles for instance. These additional sources of systematic error are included in our error estimates in \hbox{Table \ref{table1}}, indicating \vsini\ uncertainties of \hbox{$\sim 0.5$ \kms}\ and inclination uncertainties of order 10\degs. \cite{morin08v374peg} similarly estimated \vsini\ uncertainties of $\sim$1 \kms\ for the M4V star, V374 Peg.} 

The axial rotation period

\begin{equation}
\label{equation1}
P = \frac{2 \pi R_*\ {\rm sin}\ i}{v\ {\rm sin}\ i}
\end{equation}

can reliably be determined by utilising repeated starspot distortions in the line profiles for $\gtrsim$ 1 stellar rotation. We assumed uncertainties of 10\% for the estimated radii of \gj\ and \lp\ in Table 1, which are in agreement with the discrepancies between theoretical and observed M dwarf radii \citep{stassun12radii,williams15}. {Although \gj\ is considered a young disk object, which was not found to be a member of the Hyades supercluster by \cite{montes01members}, we have assumed a radius appropriate for a 600 Myr object following \cite{baraffe98}. Our radius uncertainty of 10\% corresponds to a minimum age of 100 Myr. At 3 Gyr, \cite{baraffe98} estimate a radius that is $<2$\% larger than at 600 Gyr, which is somewhat lower than our adopted 10\% uncertainty.} From the estimated radii \citep{baraffe98} in Table 1, we find maximum periods for our targets of \hbox{$P_{\rm estimated}/{\rm sin}\,i$} \hbox{= ${\bf0.4023 \pm 0.0406}$\ d} and ${0.1939 \pm 0.0200}$\ d respectively.
Using DoTS, we then performed \chisq\ minimisation searches for $P$ vs $i$ (for $P < P_{\rm estimated}$), which yielded $P = 0.3088$\ d and $i = \hbox{54}$\degs\ for \gj. Similarly for \lp\, $P = 0.1620$\ d and $i = 55$\degs\ (Figure \ref{fig1}).  That periodicities can be determined this way indicates coherence of spots on a single rotation timescale for \lp\ and on three day timescales for \gj. We note that the SuperWASP period of \hbox{0.257\ d} leads to a significantly poorer fit to the data with mis-match of repeated starspot features. The derived radii for both stars are consistent with the estimated model radii with the 10\% uncertainties.
}

\begin{figure*}
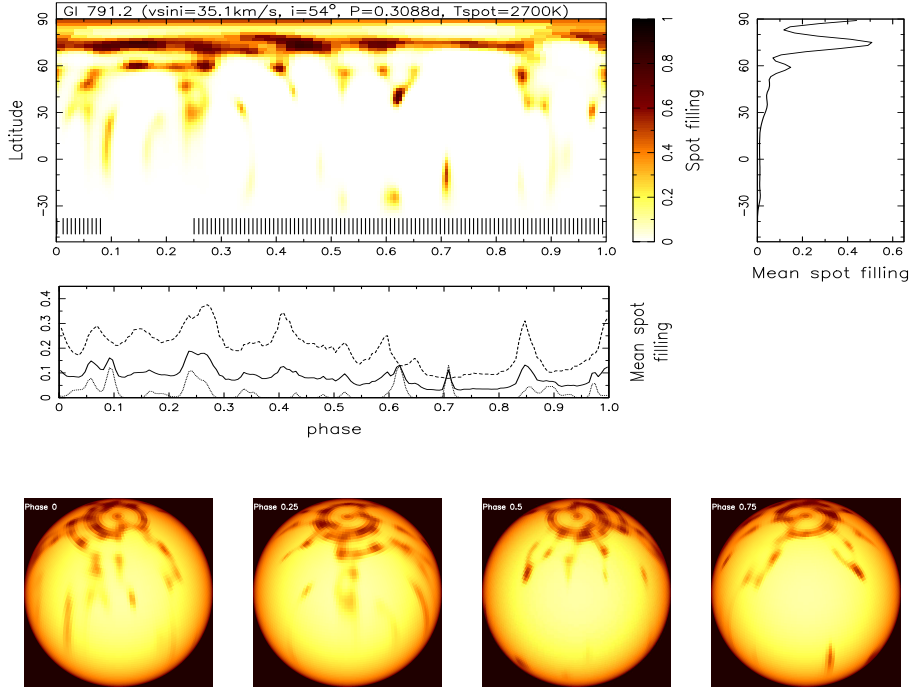

\begin{center}
\begin{tabular}{c}
\vspace{-23mm} \\
\includegraphics[width=6.4cm,height=12.0cm,angle=270,bbllx=70,bblly=20,bburx=560,bbury=765]{im_gj791_54degs_28its_2700K_aug25_label54degs.ps} \\
\end{tabular}

\begin{tabular}{cccc}
\vspace{-5mm} \\
\hspace{0mm} \includegraphics[width=2.5cm,angle=270,bbllx=40,bblly=150,bburx=516,bbury=627]{gj791_phase000-2700K.ps2} &
\hspace{0mm} \includegraphics[width=2.5cm,angle=270,bbllx=40,bblly=150,bburx=516,bbury=627]{gj791_phase250-2700K.ps2} &
\hspace{0mm} \includegraphics[width=2.5cm,angle=270,bbllx=40,bblly=150,bburx=516,bbury=627]{gj791_phase500-2700K.ps2} &
\hspace{0mm} \includegraphics[width=2.5cm,angle=270,bbllx=40,bblly=150,bburx=516,bbury=627]{gj791_phase750-2700K.ps2} \\

\vspace{-5mm} \\
\end{tabular}
\end{center}
\caption{Maximum entropy regularized image reconstructions of \gj\ plotted as Mercator (top) and 3D projections (bottom). The image of \gj\ is shown for parameters optimised with $T_{\rm phot} = 3000$\ K and $T_{\rm spot} = 2700$\ K, which have a contrast ratio at disk center of $I_{\rm spot}(0)/I_{\rm phot}(0)$ \hbox{= 0.41}. The tick marks in the Mercator plot indicate the phases of observation. The latitudinal mean spot filling is shown in the upper left panel. The mean phase spot filling is indicated in the panel below the Mercator map for latitudes -30\degs\ to 90\degs (solid line), -30\degs\ to 45\degs (dotted line) and 45\degs\ to 90\degs\ (dashed line).\label{recongj791}}
\end{figure*}

\section{Results}
\label{section:results}
{
The observed time series deconvolved profiles, along with the regularised fits using the parameters in \hbox{Table \ref{table1}} are shown for both stars in \hbox{Figure \ref{decon-timeseries-both}}. To enhance the appearance of starspots, the time series profiles are shown as 2D colorscale plots divided by the mean profile in Figure \ref{timeseriesgj791} for \gj\ and Figure \ref{timeserieslp944} for \lp. Figures \ref{timeseriesgj791} \& \ref{timeserieslp944} also show the fits with the mean profile removed along with the residual time series. The reconstructed maps in Figures \ref{recongj791} - \ref{reconlp944}, which have 2\degs\ pixel resolution, reveal spots reconstructed at a range of latitudes. The gradient of a starspot feature in the time series spectra determines the stellar latitude at which it is reconstructed, while the recovered phase is given by the time at which the feature crosses the profile center.  The right hand panels in Figures \ref{recongj791} - \ref{reconlp944} indicate the mean spot filling at each latitude, while the mean spot filling as a function of phase is shown for different latitude ranges in the panels below each Mercator projection. The 3D images at rotation phases 0.00, 0.25, 0.50 \& 0.75 are also shown for all cases.

}

\subsection{\gj}
\label{section:gj791results}
GJ 791.2A shows a number of starspot features of different sizes distributed across most phases. We used $T_{\rm phot}$ = 3000 K and $T_{\rm spot}$ \hbox{= 2700 K}, with a disk center contrast ratio of $I^c_{\rm spot}/I^c_{\rm phot}$ \hbox{= 0.41} for the image reconstruction in Figure \ref{recongj791}. The phase range \hbox{$\phi =$} \hbox{0.0000\,-} \,0.0803 is provided by observations on September 3 that align spot features with those from September 6, which first come into view at phases \hbox{$\phi \sim$} \hbox{0.8\,-\,1.0}. Phase overlap between nights is confined to the narrow \hbox{$\phi =$} \hbox{0.0000\,-\,0.0008} range, corresponding to the first observation on September 3 and the last observation on September 6.
Any starspots at the center of the 0.0803\ -\ 0.2504 phase gap in observations are still seen for 66\% of the total time they are physically visible on the star. Spots in the phase gap can thus be reconstructed, albeit with potentially decreased reliability in latitude. 


Since cos($\theta$) $\propto$ starspot trail gradient to first order (where $\theta$ is the stellar latitude), and cos($\theta$) is a slowly changing function when $\theta$ is small, the elongation of spots at low latitudes is a reflection of the uncertainty in $\theta$. There are $\sim17.5$ resolution elements in the rotationally broadened profile of \gj, equivalent to $\sim 10.3$\degs\ resolution, or $0.029$ in phase, at the equator (the 180 sec exposures result in rotational blurring of only 2.2\degs). The weakest low latitude features appear elongated with FWHM values of \hbox{$\Delta\phi$ $\sim4$\degs\,-\,$6$\degs} and \hbox{$\Delta\theta$ $\sim22$\degs\,-\,$30$\degs}. These features are approximately equivalent to circular spots with radii $\sim 9.5\,-\,13.5$\degs.

The maximum spot filling factor in the image of \gj\ is $f_{\rm max} = 0.823$ (82.3\%) and the mean spot filling is $\bar{f} = 0.032$ (3.2\%). Spot activity is seen at all latitudes, but with peak filling factors concentrated at \hbox{$\theta$ = 59\degs} and particularly \hbox{$\theta$ = 75\degs} as indicated in Figure \ref{recongj791} (panel to right of Mercator map). Spot coverage as a function of phase (Figure \ref{recongj791}, panel below Mercator map) is higher on average in the 45\degs\,-\,90\degs\ region (dashed line) compared with the lower -30\degs\,-\,45\degs\ region (dotted line), again demonstrating the greater degree of spot filling at higher latitudes. To some extent this is due to the fact that the number of resolution elements is greater at lower latitudes where features are better resolved. In other words, in the -30\degs\,-\,45\degs\ range, the spot coverage as a function of phase approaches zero in regions outside the spots, which can be seen to produce more narrow peaks than for the 45\degs\,-\,90\degs\ region.

\subsubsection{Images with $\Delta T$ = 200 K \& 400 K}
\label{section:gj791results2}

We also attempted to fit our spectra with \hbox{$T_{\rm spot}$ = 2800 K} ($\Delta T = 200$\ K) with an intensity contrast ratio at disk center of $I^c_{\rm spot}/I^c_{\rm phot}$ \hbox{= 0.57}. However, for  $T_{\rm spot}$ \hbox{= 2800 K}, we achieved a {reduced \chisq\ of \chisqr\ = 2.39}, whereas with $T_{\rm spot}$ \hbox{= 2700 K}, we achieved \chisqr\ = 1.35. The contrast in the corresponding fits is lower in Figure \ref{timeseriesgj791} (lower panels) with noticeably greater residuals. The upper panels in Figure \ref{recongj791-param} show the image for $T_{\rm spot}$ = 2800 K which shows a maximum spot filling of $f_{\rm max} = 0.983$ and a mean spot filling, $\bar{f} = 0.038$. The image is very similar to the $T_{\rm spot}$ = \hbox{2700 K} image, except for the greater degree of spot filling. As expected, DoTS has tried to account for the lack of contrast by enlarging some of the spots (e.g. the spot at phase 0.62 and -30\degs) in an attempt to fit the starspot distortions in the line profiles. 

With $T_{\rm spot}$ = 2600 K ($\Delta T =$ \hbox{$400$\ K}), we find a contrast of $I^c_{\rm spot}/I^c_{\rm phot}$ = \hbox{0.29}. The reconstructed image closely resembles the $\Delta T =$\ \hbox{$300$\ K} image, but shows reduced spot filling factors of $f_{\rm max} = 0.690$ and a mean spot filling, $\bar{f} = 0.026$ \hbox{(\chisqr\ = 1.38)}.

\subsection{The effect of limb darkening uncertainty on image reconstruction}
\label{section:gj791variability}

The concentration of spots seen on \gj\ in Figure \ref{recongj791} at 75\degs\ appear to occur in a relatively narrow band, which is most evident in the 3D plots shown below the Mercator projection in Figure \ref{recongj791}. Features that do not traverse the whole \vsini\ extent of the star are reconstructed at high latitudes. Closely separated starspot trails (in phase), or broader, more complex trails can not be resolved and can result in merging of starspot features in the images. At higher latitudes, this can lead to the appearance of arcs or rings of structure. To further investigate the axisymmetric structure at 75\degs\ on \gj\ further, we have explored a number of different input models.

\begin{figure*}
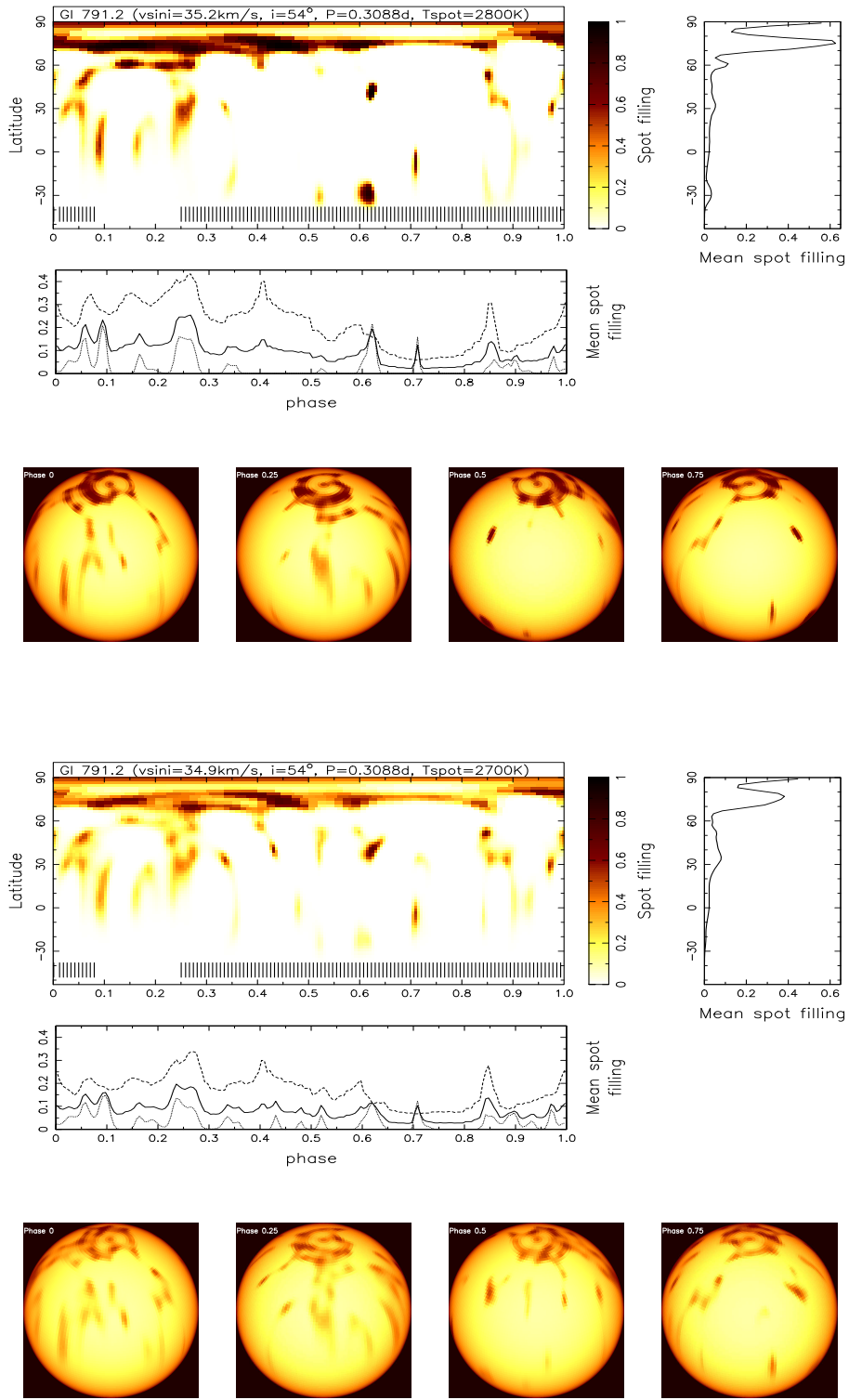

\begin{center}

\begin{tabular}{c}
\vspace{-10mm} \\
\includegraphics[width=6.4cm,height=12.0cm,angle=270,bbllx=70,bblly=20,bburx=560,bbury=765]{im_gj791_54degs_28its_2800K_aug25_label54degs.ps} \\
\end{tabular}

\begin{tabular}{cccc}
\vspace{-5mm} \\
\hspace{0mm} \includegraphics[width=2.5cm,angle=270,bbllx=40,bblly=150,bburx=516,bbury=627]{gj791_phase000-2800K.ps2} &
\hspace{0mm} \includegraphics[width=2.5cm,angle=270,bbllx=40,bblly=150,bburx=516,bbury=627]{gj791_phase250-2800K.ps2} &
\hspace{0mm} \includegraphics[width=2.5cm,angle=270,bbllx=40,bblly=150,bburx=516,bbury=627]{gj791_phase500-2800K.ps2} &
\hspace{0mm} \includegraphics[width=2.5cm,angle=270,bbllx=40,bblly=150,bburx=516,bbury=627]{gj791_phase750-2800K.ps2} \\
\vspace{-5mm} \\
\end{tabular}

\begin{tabular}{c}
\vspace{10mm} \\
\includegraphics[width=6.4cm,height=12.0cm,angle=270,bbllx=70,bblly=20,bburx=560,bbury=765]{im_gj791_54degs_28its_2700K_ldcplus10_aug25_label54degs.ps} \\
\end{tabular}

\begin{tabular}{cccc}
\vspace{-5mm} \\
\hspace{0mm} \includegraphics[width=2.5cm,angle=270,bbllx=40,bblly=150,bburx=516,bbury=627]{gj791_phase000-ldcp10-2700K.ps2} &
\hspace{0mm} \includegraphics[width=2.5cm,angle=270,bbllx=40,bblly=150,bburx=516,bbury=627]{gj791_phase250-ldcp10-2700K.ps2} &
\hspace{0mm} \includegraphics[width=2.5cm,angle=270,bbllx=40,bblly=150,bburx=516,bbury=627]{gj791_phase500-ldcp10-2700K.ps2} &
\hspace{0mm} \includegraphics[width=2.5cm,angle=270,bbllx=40,bblly=150,bburx=516,bbury=627]{gj791_phase750-ldcp10-2700K.ps2} \\
\end{tabular}

\end{center}
\caption{As in Figure \ref{recongj791} for $T_{\rm spot} = 2800$\ K (top), $T_{\rm spot} = 2700$\ K (bottom) with limb darkening reduced by 10\% with a linear centre-to-limb scaling (bottom).\label{recongj791-param}}
\end{figure*}

Overestimating the degree of limb darkening leads to a more V-shaped line profile, requiring a higher \vsini\ to fit the time series spectra. On the other hand, a lower degree of limb darkening leads to a more U-shaped profile and a typically lower \vsini\ estimate, potentially reducing circumpolar structure. To assess the reality of the high latitude axisymmetric structure on \gj, we investigated the effect of modifying the limb darkening in our input models. We scaled the center-to-limb intensities at each limb-angle using a linear relationship, such that $I_{\rm adjusted} = I_{\rm model}(1-\xi (1-\mu))$, where $\xi$ is the scale factor. We investigated $\xi =$ -0.2, -0.1, 0.1 \& 0.2 and found that 
the \chisq\ after fitting our profiles was minimised by reducing the limb darkening by 10\% (i.e. $\xi = -0.1$). A reduced \chisqr\ = 1.18 was achieved compared with \chisqr\ = 1.35 for the default case. The resulting image is shown in Figure \ref{recongj791-param} (lower panels) which shows that the axisymmetric ring at high latitude structure is less prominent. Mean and maximum spot filling factors of $\bar{f} = 0.033$ and $f_{\rm max} = 0.76$ are found.

With a 20\% reduction in limb darkening, the high latitude axisymmetric ring structure is reduced in prominence, leaving only the main concentrations of spots, but all lower latitude structure becomes confined to the latitude range 0\degs\,-\,50\degs (\chisqr\ = 1.38 is achieved). Increasing the degree of limb darkening led to strong polar filling factors and poorer fits. We also investigated using $T_{\rm phot}$ = 3100 K with \hbox{$\Delta T = 400$\ K} (a contrast ratio of $I^c_{\rm spot}/I^c_{\rm phot}$ \hbox{= 0.45}). In this case, the center-to-limb intensity variation was similar to our optimal limb darkening (i.e. when reduced by 10\%) for $T_{\rm phot}$ \hbox{= 3000 K}. The center-to-limb trend is however more linear and leads to an augmentation of the axisymmetric ring compared with the image in Figure \ref{recongj791} with a reduced \chisqr\ = 2.04. 

Finally, removing the center-to-limb line equivalent width variation had little discernible effect on the image, although the strength of the spot filling at 75\degs\ increased by 9\%. Optimising the center-to-limb variation in intensity and the non-linearity of the limb darkening is thus important for ensuring that artefacts are minimised. The need to reduce the limb darkening by 10\% gives an empirical indication of the systematic uncertainty in the current models for cool M dwarfs. 

\subsection{LP 944-20}
\label{section:lp944results}

{ 
The two dominant starspot trails seen in the time series spectra of \lp\ (Figure \ref{timeserieslp944}) possess steep gradients, leading us to expect starspots should be reconstructed at predominantly high latitudes. The image of \lp\ in Figure \ref{reconlp944} (top two panels) is reconstructed with \hbox{$i$ = 55\degs}\ and \hbox{$P =$} \hbox{$0.1620$ d} and shows activity is concentrated in two groups at $\phi \sim 0.2$ \& $0.65$, approximately half a rotation phase apart. For \vsini\ = \hbox{30.8 \kms}, the equatorial spot resolution is 11.5\degs. As with \gj, spots are found at high latitude, with a maximum in the azimuthally averaged spot filling at 79\degs. The spot reconstructed at $\phi = 0.16$ and $\theta = 55$\degs\ is predominantly responsible for the mean latitude filling peak at $\theta = 55$\degs. Given the lower S/N of the data, the smaller, weaker spots at $\phi = 0.65$, $\theta = 55$\degs\ and $\phi = 0.925$, $\theta = 35$\degs\ may be artefacts in the reconstruction. We also note that if lower latitude spots of similar size to those seen on \gj\ are present on \lp, they may not be recovered owing to the lower SNR of the data. \lp\ reveals a very low mean spot filling factor of $\bar{f} = 0.015$ (1.5\%) for the image reconstructed with $T_{\rm phot}$ = 2300 K and $T_{\rm spot}$ \hbox{= 2100 K} \hbox{($\Delta T = 200$ K)}, corresponding to a contrast ratio at disk center, $I^c_{\rm spot}/I^c_{\rm phot}$ \hbox{= 0.48}. A maximum spot filling factor, $f_{\rm max} = 0.766$ (76.6\%), is found and we achieve \chisqr\ = 1.17. 

\subsubsection{\lp\ image with $\Delta T$ = 100 K}
\label{section:lp944results2}

The image reconstruction with $T_{\rm spot}$ \hbox{= 2200 K} ($\Delta T = 100$\ K, $I^c_{\rm spot}/I^c_{\rm phot}$ \hbox{= 0.64}) is shown in Figure \ref{reconlp944} (bottom two panels). As with \gj, when we investigated a reduced contrast, we find that the data are not fit so well (\chisqr\ = 1.22) {compared with $T_{\rm spot}$ \hbox{= 2100 K}}, and that features are reconstructed with larger areas to account for the lower contrast ratio. As expected, the mean and maximum spot filling values are higher than for ($\Delta T = 200$K), with $\bar{f} = 0.017$ and $f_{\rm max} = 0.866$. The lower SNR of the \lp\ time series makes investigation of limb darkening modification more difficult. However, we found that even for $\xi = 0.2$ (20\% {\em increase} in limb darkening), a \chisqr\ = 1.17 is achieved. The image shows the same features as for our default $\Delta T = 200$ K, but with enhanced circum-polar filling, as might be expected by increasing the limb darkening ($\bar{f} = 0.019$, $f_{\rm max} = 0.88$).

}



\section{RV variability}
\label{section:rvjitter}

We have cross-correlated the observed time series to determine the RV variability induced by the spots on both targets. We find a semi-amplitude of { \hbox{$K_{*} =$ 160\ \ms}\ for \gj, with a corresponding RV r.m.s. of \hbox{138\ \ms}. We also cross-correlated the time series line profile fits for $\Delta T$ = \hbox{300 K}. {By subtracting the RVs derived from the fits from those measured for the data}, following \cite{donati14lkca4}, the starspot induced radial velocity variability in the data can be partially corrected. We find that the radial velocity r.m.s. can be reduced by a factor of 1.75, from \hbox{138\ \ms}\ to \hbox{79 \ms}. Using the optimised fits with $\Delta T$ = \hbox{300 K}\ and the limb darkening reduced by 10\%, we find the RV r.m.s. is reduced by a factor of 1.9, to 73 \ms. For LP944-20, the SNR is not sufficient to discern regular RV variability. The radial velocity r.m.s. of 194 \ms\ is thus likely to be noise dominated. The regularised fits however yield an RV r.m.s value of \hbox{93 \ms}.

\citet{barnes11jitter} found similar RV r.m.s. values to those found here for \gj\ and \lp\ for their scaled solar starspot model with \hbox{\vsini\ =} \hbox{20\,-\,50 \kms}. However, models with large spot filling fractions of either $f = 0.30$ and greater spot-to-photosphere contrast than found here (up to $T_{\rm phot}$/$T_{\rm spot}$ = 0.65) or models with $f = 0.62$ and similar contrast ($\Delta T = 250$K) are required to obtain r.m.s.\ RVs of order \hbox{100 \ms}. {Based on the evidence presented here, it thus appears spots or spot groups are larger or more localised than the random scaled solar spots models of \citet{barnes11jitter}.}

\begin{figure*}
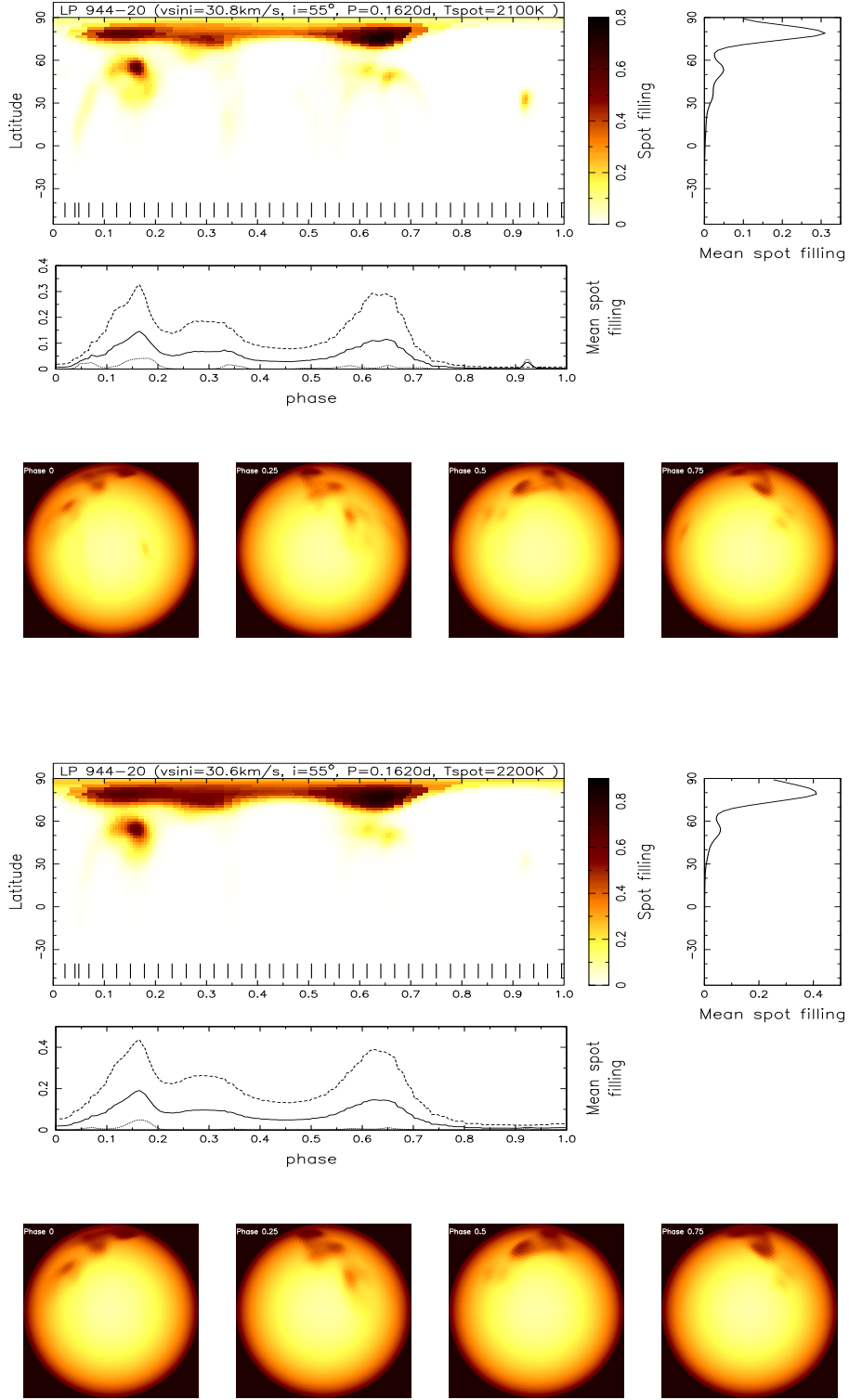

\begin{center}

\begin{tabular}{c}
  \vspace{-23mm} \\
  \includegraphics[width=6.4cm,height=12.0cm,angle=270,bbllx=70,bblly=20,bburx=560,bbury=765]{im_lp944_55degs_19its_2100K_jun03.ps} \\
\end{tabular}

\begin{tabular}{cccc}
  \vspace{-5mm} \\
  \hspace{0mm} \includegraphics[width=2.5cm,angle=270,bbllx=40,bblly=150,bburx=516,bbury=627]{lp944_phase000_jun03.ps2} &
  \hspace{0mm} \includegraphics[width=2.5cm,angle=270,bbllx=40,bblly=150,bburx=516,bbury=627]{lp944_phase250_jun03.ps2} &
  \hspace{0mm} \includegraphics[width=2.5cm,angle=270,bbllx=40,bblly=150,bburx=516,bbury=627]{lp944_phase500_jun03.ps2} &
  \hspace{0mm} \includegraphics[width=2.5cm,angle=270,bbllx=40,bblly=150,bburx=516,bbury=627]{lp944_phase750_jun03.ps2} \\
\end{tabular}

\begin{tabular}{c}
  \vspace{10mm} \\
  \includegraphics[width=6.4cm,height=12.0cm,angle=270,bbllx=70,bblly=20,bburx=560,bbury=765]{im_lp944_55degs_19its_2200K_jun03.ps} \\
\end{tabular}

\begin{tabular}{cccc}
  \vspace{-5mm} \\
  \hspace{0mm} \includegraphics[width=2.5cm,angle=270,bbllx=40,bblly=150,bburx=516,bbury=627]{lp944_phase000_jun03_2p2K.ps2} &
  \hspace{0mm} \includegraphics[width=2.5cm,angle=270,bbllx=40,bblly=150,bburx=516,bbury=627]{lp944_phase250_jun03_2p2K.ps2} &
  \hspace{0mm} \includegraphics[width=2.5cm,angle=270,bbllx=40,bblly=150,bburx=516,bbury=627]{lp944_phase500_jun03_2p2K.ps2} &
  \hspace{0mm} \includegraphics[width=2.5cm,angle=270,bbllx=40,bblly=150,bburx=516,bbury=627]{lp944_phase750_jun03_2p2K.ps2} \\
\end{tabular}

\end{center}
\caption{\bf{Maximum entropy regularized image reconstructions of \lp\ with panels as indicated in Figure \ref{recongj791}. Images are shown for $T_{\rm phot} = 2300$\ K, with $T_{\rm spot} = 2100$\ K (upper panels) and $T_{\rm spot} =$ \hbox{$2200$\ K} (lower panels), with respective contrast ratios of $I_{\rm spot}(0)/I_{\rm phot}(0)$ \hbox{= 0.48} \& $0.64$. \label{reconlp944}}}
\end{figure*}

\section{Discussion}
\label{section:discussion}

The time series spectra and Doppler images of the two mid-late M dwarfs studied here reveal significant spot structure, even at spectral type M9V. {The results are in general agreement with the finding that the fall-off in fractional \ha\ flux in late M dwarfs \citep{mohanty03activity} is not in fact accompanied by a drop-off in magnetic flux \citep{reiners07magnetic}, and provide further evidence for magnetic activity in the lowest mass M dwarfs.}
The image of \gj\ shows starspots located at a range of latitudes and longitudes, but preferentially at mid to high latitudes, whereas activity is confined solely to high latitudes on \lp. Late-F and early-G dwarfs, show a still greater degree of spot filling at high latitudes compared with lower latitudes, with activity confined to localised regions in phase at lower latitude (e.g. \citealt{barnes01aper,marsden06hd171488}). {In contrast}, Doppler images of the M1-2V stars, \hbox{HK Aqr} and \hbox{EY Dra} \citep{barnes01mdwarfs,barnes04hkaqr} showed distributed activity, while differences are also seen when compared with images of the M4V stars \hbox{V394 Peg} \citep{morin08v374peg} and \hbox{G164-31} \citep{phanbao09} which possess similar \vs\ values (\hbox{36.5\ \kms}\ and \hbox{41\ \kms}\ respectively). Morin et al. found weak features concentrated at mid latitudes with only 2\% mean spot occupancy, similar to, but slightly lower than the 3.3\% we find for the optimised image of \gj. Phan-Bao et al. {on the other hand} found only a weak de-centered polar spot with no low-latitude features on \hbox{G164-31}. The current evidence from Doppler images thus suggests that the lower photometric variability in cooler M dwarfs is the result of both lower spot contrast and more distributed activity {at predominantly higher latitudes}.


{Zeeman Doppler imaging does not reveal the field strength in the largest spots, which show significant spot/photosphere contrast, because the amplitude of the \hbox{Stokes V} signature is generally small. Comparing magnetic maps and brightness maps is often difficult as a result, and because \hbox{Stokes V} only sees the {\em net} magnetic field, thus making it sensitive to field structure on larger scales. Large} scale poloidal fields were inferred for \hbox{V394 Peg} \citep{morin08v374peg} and \hbox{G164-31} \cite{phanbao09}, {in agreement with the results for more slowly rotating fully convective stars \citep{morin08mdwarfs}}. {Both these stars} showed only positive magnetic polarity in one hemisphere. These studies show contradictory results when the magnetic and brightness images are compared, since \hbox{V394 Peg} {possessed} multiple low to intermediate latitude features, {and no polar features}, while \hbox{G164-31} {showed only} a single weak circum-polar feature. {This probably suggests that even for M4V stars, the spot/photosphere contrast ratios are sufficient to preclude the detection of \hbox{Stokes V} signatures in the spots, meaning that magnetic and brightness maps indeed probe magnetic activity on different scales.} 


{Our images show a number of polar or circum-polar spots, especially in \gj\ that might reasonably be expected to possess a mixture of magnetic polarities. This is interesting because the degree to which the polarities cancel will likely determine the large scale global field structure, which in fully convective mid-M stars is generally axisymmetric and poloidal \citep{morin08mdwarfs,morin08v374peg,phanbao09}. Although we have already noted that spots recovered in brightness images do not generally correlate with magnetic field structure recovered from Stokes V, the lower contrasts in \gj\ and \lp, with only \hbox{T$_{\rm phot}$ - T$_{\rm spot}$} \hbox{= 200\,-\,300 K} ($I^c_{\rm spot}/I^c_{\rm phot}$ \hbox{$\sim 0.4\,-\,0.5$}), may enable at least some of the field in the spots to be more readily seen in Stokes V. This will also depend on the mixing and degree of polarity cancelling on relatively small scales and the resolution limitations of any observations.

With low contrast spots, future \hbox{Stokes V} observations of M dwarfs may provide the means to make more direct comparisons between spot and magnetic field morphologies and to determine why the fraction of magnetic flux seen in \hbox{Stokes V} compared with \hbox{Stokes I} is greater than for early M spectral types \citep{reiners09topology}. {In particular, infrared polarimetric observations may offer the potential to observe the magnetic field inside starspots since the contrast ratio between spots and photosphere is smaller at longer wavelengths}. We also note that \citet{hallinan15nature} have suggested a transition from coronal activity to auroral activity in late M stars and have suggested that such activity arises from circumpolar latitudes on the M8.5V, LSR J1835. It is not clear whether the high latitude, low contrast cool spots recovered in \lp\ could be related to, or contradict this mechanism.




}

Dedicated RV surveys such as the Habitable Zone Planet Finder \citep{mahadevan10hzp}, CARMENES \citep{quirrenbach10} and SPIRou \citep{thibault12spirou} will search for low mass planets around M dwarfs. Consequently, the direct influence of magnetic activity on stellar lines is of significant importance. For stars rotating with \hbox{\vsini\ = 20 \kms}, \citet{barnes11jitter} found that \hbox{5\ \mearth}\ to \hbox{20\ \mearth}\ habitable zone planets could be detected around \hbox{0.1 \msun}, M6V stars, but would require 300 and 40 epochs respectively. Planets with \hbox{1\mearth}\ could nevertheless be detected with $\sim100$ epochs for stars with \hbox{\vsini\ =} \hbox{5 \kms} and (\hbox{$T_{\rm phot} - T_{\rm spot}$ =} \hbox{$250$\ K}). 
These estimates will likely be modified by the findings presented here, since we find spot groups are more localised and larger (especially in the case of \lp) than the spots modeled by \citet{barnes11jitter}.

\citet{martin06} have searched for RV variability in LP944-20. In the optical, 15 observations spanning 841 days, at similar instrumental setup to the observations presented herein, yielded a semi-amplitude variability of \hbox{3.5\ \kms}. By contrast, Keck/NIRSPEC observations spanning only 6 nights yielded a dispersion of \hbox{360 \ms}; an order of magnitude smaller than at optical wavelengths. 
Our finding of \hbox{194\ \ms} RV r.m.s. is thus closer in magnitude to the NIRSPEC observations. Significant spot-coverage evolution, as evidenced through other activity indicators in M dwarfs \citep{gomes11mdwarfs}, is thus likely to have biased the results of \citet{martin06}. Given the observed activity in M dwarfs, it is clear that searches for low mass planets will benefit from techniques that can effectively reduce or remove the substantial RV noise. Although one might not routinely expect to observe stars with rotation velocities of order 30 \kms, the fact that Doppler imaging can be used to mitigate the effects of spots on \gj\ by a factor of $\sim 2$, indicates that it may also be a useful method for slower rotators. This is particularly important in mid-late M stars, since they show equatorial rotation velocities of $\sim10$ \kms\ on average \citep{jenkins09mdwarfs}. Observing only the very slowest rotators is likely to bias the sample to low axial inclinations, for which it would be much more difficult to detect orbiting planets.

\section{Summary}
\label{section:summary}

{Significant starspot activity is seen on M stars at the bottom of the main sequence, albeit at lower contrast compared with G and K dwarfs. Using model atmospheres as input for our image reconstructions, we find that $\Delta T$ \hbox{$\gtrsim 300$ K} (contrast at disk center,  $I^c_{\rm spot}/I^c_{\rm phot}$ \hbox{= 0.41}) is required to model spots on the M4.5V, \gj, whereas  the data can be fit with $\Delta T$ = \hbox{100K\,-\,200\ K} (contrast at disk center,  $I^c_{\rm spot}/I^c_{\rm phot}$ \hbox{= 0.64\,-\,0.48} for the M9V \lp, albeit with a marginally better fit for $\Delta T$ = 200K. The factor of two lower spot filling for \lp\ ($f = 0.015$) compared with \gj\ ($f=0.033$) suggests a decline in magnetic activity by M9V, as seen from other magnetic activity indicators \citep{mohanty03activity}.  However, Doppler imaging at optical wavelengths is extremely challenging for even the brightest M9V targets and it is possible that smaller, low latitude features have not been recovered on \lp, owing to the lower SNR of the data. Our ability to reconstruct small scale brightness features at all latitudes should benefit from the next generation of instruments offering higher throughput and optical \citep{zerbi14hires} to near infrared wavelength coverage in conjunction with near infrared polarimetric measurements.}. 

\acknowledgments
We are grateful to the referee for providing careful and thorough feedback during this work. We would like to thank Prof. Ansgar Reiners for helpful discussion during this project. J.R.B. and C.A.H. and M.E.L. were supported by the STFC under the grant ST/L000776/1. S.V.J. acknowledges research funding by the Deutsche Forschungsgemeinschaft (DFG) under grant SFB 963/1, project A16. JSJ acknowledges funding by Fondecyt through grant 3110004 and partial support from CATA-Basal (PB06, Conicyt), the GEMINI-CONICYT FUND and from the Comit\'{e} Mixto ESO-GOBIERNO DE CHILE. 



\begin{thebibliography}{}
\expandafter\ifx\csname natexlab\endcsname\relax\def\natexlab#1{#1}\fi

\bibitem[{{Allard} {et~al.}(2012){Allard}, {Homeier}, {Freytag}, \&
  {Sharp}}]{allard12}
{Allard}, F., {Homeier}, D., {Freytag}, B., \& {Sharp}, C.~M. 2012, in EAS
  Publications Series, Vol.~57, EAS Publications Series, ed. C.~{Reyl{\'e}},
  C.~{Charbonnel}, \& M.~{Schultheis}, 3--43

\bibitem[{{Baraffe} {et~al.}(1998){Baraffe}, {Chabrier}, {Allard}, \&
  {Hauschildt}}]{baraffe98}
{Baraffe}, I., {Chabrier}, G., {Allard}, F., \& {Hauschildt}, P.~H. 1998, A\&A,
  337, 403

\bibitem[{{Barnes} \& {Collier Cameron}(2001)}]{barnes01mdwarfs}
{Barnes}, J.~R., \& {Collier Cameron}, A. 2001, MNRAS, 326, 950

\bibitem[{{Barnes} {et~al.}(2000){Barnes}, {Collier Cameron}, {James}, \&
  {Donati}}]{barnes00pztel}
{Barnes}, J.~R., {Collier Cameron}, A., {James}, D.~J., \& {Donati}, J.-F.
  2000, MNRAS, 314, 162

\bibitem[{{Barnes} {et~al.}(2001){Barnes}, {Collier Cameron}, {James}, \&
  {Steeghs}}]{barnes01aper}
{Barnes}, J.~R., {Collier Cameron}, A., {James}, D.~J., \& {Steeghs}, D. 2001,
  MNRAS, 326, 1057

\bibitem[{{Barnes} {et~al.}(1998){Barnes}, {Collier Cameron}, {Unruh},
  {Donati}, \& {Hussain}}]{barnes98aper}
{Barnes}, J.~R., {Collier Cameron}, A., {Unruh}, Y.~C., {Donati}, J.~F., \&
  {Hussain}, G.~A.~J. 1998, MNRAS, 299, 904

\bibitem[{{Barnes} {et~al.}(2004){Barnes}, {James}, \&
  {Cameron}}]{barnes04hkaqr}
{Barnes}, J.~R., {James}, D.~J., \& {Cameron}, A.~C. 2004, MNRAS, 352, 589

\bibitem[{{Barnes} {et~al.}(2011){Barnes}, {Jeffers}, \&
  {Jones}}]{barnes11jitter}
{Barnes}, J.~R., {Jeffers}, S.~V., \& {Jones}, H.~R.~A. 2011, MNRAS, 412, 1599

\bibitem[{{Barnes} {et~al.}(2012){Barnes}, {Jenkins}, {Jones}, {Rojo},
  {Arriagada}, {Jord{\'a}n}, {Minniti}, {Tuomi}, {Jeffers}, \&
  {Pinfield}}]{barnes12rops}
{Barnes}, J.~R., {Jenkins}, J.~S., {Jones}, H.~R.~A., {et~al.} 2012, MNRAS,
  424, 591

\bibitem[{{Barnes} {et~al.}(2014){Barnes}, {Jenkins}, {Jones}, {Jeffers},
  {Rojo}, {Arriagada}, {Jord{\'a}n}, {Minniti}, {Tuomi}, {Pinfield}, \&
  {Anglada-Escud{\'e}}}]{barnes14rops}
---. 2014, MNRAS, 439, 3094

\bibitem[{{Batalha} {et~al.}(2013){Batalha}, {Rowe}, {Bryson}, {Barclay},
  {Burke}, {Caldwell}, {Christiansen}, {Mullally}, {Thompson}, {Brown},
  {Dupree}, {Fabrycky}, {Ford}, {Fortney}, {Gilliland}, {Isaacson}, {Latham},
  {Marcy}, {Quinn}, {Ragozzine}, {Shporer}, {Borucki}, {Ciardi}, {Gautier},
  {Haas}, {Jenkins}, {Koch}, {Lissauer}, {Rapin}, {Basri}, {Boss}, {Buchhave},
  {Carter}, {Charbonneau}, {Christensen-Dalsgaard}, {Clarke}, {Cochran},
  {Demory}, {Desert}, {Devore}, {Doyle}, {Esquerdo}, {Everett}, {Fressin},
  {Geary}, {Girouard}, {Gould}, {Hall}, {Holman}, {Howard}, {Howell},
  {Ibrahim}, {Kinemuchi}, {Kjeldsen}, {Klaus}, {Li}, {Lucas}, {Meibom},
  {Morris}, {Pr{\v s}a}, {Quintana}, {Sanderfer}, {Sasselov}, {Seader},
  {Smith}, {Steffen}, {Still}, {Stumpe}, {Tarter}, {Tenenbaum}, {Torres},
  {Twicken}, {Uddin}, {Van Cleve}, {Walkowicz}, \& {Welsh}}]{batalha13kepler}
{Batalha}, N.~M., {Rowe}, J.~F., {Bryson}, S.~T., {et~al.} 2013, ApJS, 204, 24

\bibitem[{{Bean} {et~al.}(2010){Bean}, {Seifahrt}, {Hartman}, {Nilsson},
  {Wiedemann}, {Reiners}, {Dreizler}, \& {Henry}}]{bean10b}
{Bean}, J.~L., {Seifahrt}, A., {Hartman}, H., {et~al.} 2010, ApJ, 713, 410

\bibitem[{{Benedict} {et~al.}(2000){Benedict}, {McArthur}, {Franz},
  {Wasserman}, \& {Henry}}]{benedict00gj791}
{Benedict}, G.~F., {McArthur}, B.~E., {Franz}, O.~G., {Wasserman}, L.~H., \&
  {Henry}, T.~J. 2000, AJ, 120, 1106

\bibitem[{{Berdyugina}(2005)}]{berdyugina05starspots}
{Berdyugina}, S.~V. 2005, Living Reviews in Solar Physics, 2, 8

\bibitem[{{Berger} {et~al.}(2001){Berger}, {Ball}, {Becker}, {Clarke}, {Frail},
  {Fukuda}, {Hoffman}, {Mellon}, {Momjian}, {Murphy}, {Teng}, {Woodruff},
  {Zauderer}, \& {Zavala}}]{berger01lp944}
{Berger}, E., {Ball}, S., {Becker}, K.~M., {et~al.} 2001, in Bulletin of the
  American Astronomical Society, Vol.~33, American Astronomical Society Meeting
  Abstracts \#198, 891

\bibitem[{{Bonfils} {et~al.}(2013){Bonfils}, {Delfosse}, {Udry}, {Forveille},
  {Mayor}, {Perrier}, {Bouchy}, {Gillon}, {Lovis}, {Pepe}, {Queloz}, {Santos},
  {S{\'e}gransan}, \& {Bertaux}}]{bonfils13mdwarfs}
{Bonfils}, X., {Delfosse}, X., {Udry}, S., {et~al.} 2013, A\&A, 549, A109

\bibitem[{{Borucki} {et~al.}(2011){Borucki}, {Koch}, {Basri}, {Batalha},
  {Brown}, {Bryson}, {Caldwell}, {Christensen-Dalsgaard}, {Cochran}, {DeVore},
  {Dunham}, {Gautier}, {Geary}, {Gilliland}, {Gould}, {Howell}, {Jenkins},
  {Latham}, {Lissauer}, {Marcy}, {Rowe}, {Sasselov}, {Boss}, {Charbonneau},
  {Ciardi}, {Doyle}, {Dupree}, {Ford}, {Fortney}, {Holman}, {Seager},
  {Steffen}, {Tarter}, {Welsh}, {Allen}, {Buchhave}, {Christiansen}, {Clarke},
  {Das}, {D{\'e}sert}, {Endl}, {Fabrycky}, {Fressin}, {Haas}, {Horch},
  {Howard}, {Isaacson}, {Kjeldsen}, {Kolodziejczak}, {Kulesa}, {Li}, {Lucas},
  {Machalek}, {McCarthy}, {MacQueen}, {Meibom}, {Miquel}, {Prsa}, {Quinn},
  {Quintana}, {Ragozzine}, {Sherry}, {Shporer}, {Tenenbaum}, {Torres},
  {Twicken}, {Van Cleve}, {Walkowicz}, {Witteborn}, \&
  {Still}}]{borucki11kepler}
{Borucki}, W.~J., {Koch}, D.~G., {Basri}, G., {et~al.} 2011, ApJ, 736, 19

\bibitem[{{Butters} {et~al.}(2010){Butters}, {West}, {Anderson}, {Collier
  Cameron}, {Clarkson}, {Enoch}, {Haswell}, {Hellier}, {Horne}, {Joshi},
  {Kane}, {Lister}, {Maxted}, {Parley}, {Pollacco}, {Smalley}, {Street},
  {Todd}, {Wheatley}, \& {Wilson}}]{butters10wasp}
{Butters}, O.~W., {West}, R.~G., {Anderson}, D.~R., {et~al.} 2010, A\&A, 520,
  L10

\bibitem[{{Claret}(2000)}]{claret00ldc4}
{Claret}, A. 2000, A\&A, 363, 1081

\bibitem[{Collier~Cameron(2001)}]{cameron01mapping}
Collier~Cameron, A. 2001, in Astrotomography - Indirect Imaging Methods in
  Observational Astronomy, ed. {{Boffin}, H.~M.~J. and {Steeghs}, D. and
  {Cuypers}, J.} (Springer (Lecture Notes in Physics)), 183--206

\bibitem[{{Crossfield} {et~al.}(2011){Crossfield}, {Barman}, \&
  {Hansen}}]{crossfield11gj1214b}
{Crossfield}, I.~J.~M., {Barman}, T., \& {Hansen}, B.~M.~S. 2011, ApJ, 736, 132

\bibitem[{{Crossfield} {et~al.}(2014){Crossfield}, {Biller}, {Schlieder},
  {Deacon}, {Bonnefoy}, {Homeier}, {Allard}, {Buenzli}, {Henning}, {Brandner},
  {Goldman}, \& {Kopytova}}]{crossfield14browndwarf}
{Crossfield}, I.~J.~M., {Biller}, B., {Schlieder}, J.~E., {et~al.} 2014,
  Nature, 505, 654

\bibitem[{{Delfosse} {et~al.}(1998){Delfosse}, {Forveille}, {Perrier}, \&
  {Mayor}}]{delfosse98mdwarfs}
{Delfosse}, X., {Forveille}, T., {Perrier}, C., \& {Mayor}, M. 1998, A\&A, 331,
  581

\bibitem[{{Dieterich} {et~al.}(2014){Dieterich}, {Henry}, {Jao}, {Winters},
  {Hosey}, {Riedel}, \& {Subasavage}}]{dieterich14}
{Dieterich}, S.~B., {Henry}, T.~J., {Jao}, W.-C., {et~al.} 2014, AJ, 147, 94

\bibitem[{{Donati} {et~al.}(2008){Donati}, {Morin}, {Petit}, {Delfosse},
  {Forveille}, {Auri{\`e}re}, {Cabanac}, {Dintrans}, {Fares}, {Gastine},
  {Jardine}, {Ligni{\`e}res}, {Paletou}, {Velez}, \&
  {Th{\'e}ado}}]{donati08mdwarfs}
{Donati}, J., {Morin}, J., {Petit}, P., {et~al.} 2008, MNRAS, 390, 545

\bibitem[{{Donati} {et~al.}(2014){Donati}, {H{\'e}brard}, {Hussain}, {Moutou},
  {Grankin}, {Boisse}, {Morin}, {Gregory}, {Vidotto}, {Bouvier}, {Alencar},
  {Delfosse}, {Doyon}, {Takami}, {Jardine}, {Fares}, {Cameron}, {M{\'e}nard},
  {Dougados}, {Herczeg}, \& {Matysse Collaboration}}]{donati14lkca4}
{Donati}, J.-F., {H{\'e}brard}, E., {Hussain}, G., {et~al.} 2014, MNRAS, 444,
  3220

\bibitem[{{Gadun} \& {Pavlenko}(1997)}]{gadun97}
{Gadun}, A.~S., \& {Pavlenko}, Y.~V. 1997, \aap, 324, 281

\bibitem[{{Gomes da Silva} {et~al.}(2011){Gomes da Silva}, {Santos}, {Bonfils},
  {Delfosse}, {Forveille}, \& {Udry}}]{gomes11mdwarfs}
{Gomes da Silva}, J., {Santos}, N.~C., {Bonfils}, X., {et~al.} 2011, A\&A, 534,
  A30

\bibitem[{{Hallinan} {et~al.}(2015){Hallinan}, {Littlefair}, {Cotter},
  {Bourke}, {Harding}, {Pineda}, {Butler}, {Golden}, {Basri}, {Doyle}, {Kao},
  {Berdyugina}, {Kuznetsov}, {Rupen}, \& {Antonova}}]{hallinan15nature}
{Hallinan}, G., {Littlefair}, S., {Cotter}, G., {et~al.} 2015, Nature, 523, 568

\bibitem[{Horne(1986)}]{horne86extopt}
Horne, K.~D. 1986, PASP, 98, 609

\bibitem[{{Jenkins} {et~al.}(2009){Jenkins}, {Ramsey}, {Jones}, {Pavlenko},
  {Gallardo}, {Barnes}, \& {Pinfield}}]{jenkins09mdwarfs}
{Jenkins}, J.~S., {Ramsey}, L.~W., {Jones}, H.~R.~A., {et~al.} 2009, ApJ, 704,
  975

\bibitem[{{Kaltenegger} \& {Traub}(2009)}]{kalteneggar09}
{Kaltenegger}, L., \& {Traub}, W.~A. 2009, ApJ, 698, 519

\bibitem[{{Knutson} {et~al.}(2011){Knutson}, {Madhusudhan}, {Cowan},
  {Christiansen}, {Agol}, {Deming}, {D{\'e}sert}, {Charbonneau}, {Henry},
  {Homeier}, {Langton}, {Laughlin}, \& {Seager}}]{knutson11gj436b}
{Knutson}, H.~A., {Madhusudhan}, N., {Cowan}, N.~B., {et~al.} 2011, ApJ, 735,
  27

\bibitem[{{Kupka} {et~al.}(1999){Kupka}, {Piskunov}, {Ryabchikova}, {Stempels},
  \& {Weiss}}]{kupka99}
{Kupka}, F., {Piskunov}, N., {Ryabchikova}, T.~A., {Stempels}, H.~C., \&
  {Weiss}, W.~W. 1999, A\&AS, 138, 119

\bibitem[{{Lohr} {et~al.}(2014){Lohr}, {Norton}, {Anderson}, {Collier Cameron},
  {Faedi}, {Haswell}, {Hellier}, {Hodgkin}, {Horne}, {Kolb}, {Maxted},
  {Pollacco}, {Skillen}, {Smalley}, {West}, \& {Wheatley}}]{lohr14}
{Lohr}, M.~E., {Norton}, A.~J., {Anderson}, D.~R., {et~al.} 2014, A\&A, 566,
  A128

\bibitem[{{Mahadevan} {et~al.}(2010){Mahadevan}, {Ramsey}, {Wright}, {Endl},
  {Redman}, {Bender}, {Roy}, {Zonak}, {Troupe}, {Engel}, {Sigurdsson},
  {Wolszczan}, \& {Zhao}}]{mahadevan10hzp}
{Mahadevan}, S., {Ramsey}, L., {Wright}, J., {et~al.} 2010, in SPIE, Vol. 7735,
  SPIE, 6

\bibitem[{{Marsden} {et~al.}(2006){Marsden}, {Donati}, {Semel}, {Petit}, \&
  {Carter}}]{marsden06hd171488}
{Marsden}, S.~C., {Donati}, J.-F., {Semel}, M., {Petit}, P., \& {Carter}, B.~D.
  2006, MNRAS, 370, 468

\bibitem[{{Mart{\'{\i}}n} {et~al.}(2006){Mart{\'{\i}}n}, {Guenther}, {Zapatero
  Osorio}, {Bouy}, \& {Wainscoat}}]{martin06}
{Mart{\'{\i}}n}, E.~L., {Guenther}, E., {Zapatero Osorio}, M.~R., {Bouy}, H.,
  \& {Wainscoat}, R. 2006, ApJ, 644, L75

\bibitem[{{Mohanty} \& {Basri}(2003)}]{mohanty03activity}
{Mohanty}, S., \& {Basri}, G. 2003, ApJ, 583, 451

\bibitem[{{Montes} {et~al.}(2001){Montes}, {L{\'o}pez-Santiago}, {G{\'a}lvez},
  {Fern{\'a}ndez-Figueroa}, {De Castro}, \& {Cornide}}]{montes01members}
{Montes}, D., {L{\'o}pez-Santiago}, J., {G{\'a}lvez}, M.~C., {et~al.} 2001,
  MNRAS, 328, 45

\bibitem[{{Morin} {et~al.}(2010){Morin}, {Donati}, {Petit}, {Delfosse},
  {Forveille}, \& {Jardine}}]{morin10mdwarfs}
{Morin}, J., {Donati}, J.-F., {Petit}, P., {et~al.} 2010, MNRAS, 407, 2269

\bibitem[{{Morin} {et~al.}(2008{\natexlab{a}}){Morin}, {Donati}, {Petit},
  {Delfosse}, {Forveille}, {Albert}, {Auri{\`e}re}, {Cabanac}, {Dintrans},
  {Fares}, {Gastine}, {Jardine}, {Ligni{\`e}res}, {Paletou}, {Ramirez Velez},
  \& {Th{\'e}ado}}]{morin08mdwarfs}
{Morin}, J., {Donati}, J., {Petit}, P., {et~al.} 2008{\natexlab{a}}, MNRAS,
  390, 567

\bibitem[{{Morin} {et~al.}(2008{\natexlab{b}}){Morin}, {Donati}, {Forveille},
  {Delfosse}, {Dobler}, {Petit}, {Jardine}, {Collier Cameron}, {Albert},
  {Manset}, {Dintrans}, {Chabrier}, \& {Valenti}}]{morin08v374peg}
{Morin}, J., {Donati}, J.-F., {Forveille}, T., {et~al.} 2008{\natexlab{b}},
  MNRAS, 384, 77

\bibitem[{{Pavlenko}(2014)}]{pavlenko14}
{Pavlenko}, Y.~V. 2014, Astronomy Reports, 58, 825

\bibitem[{{Pavlenko} {et~al.}(2007){Pavlenko}, {Jones}, {Mart{\'{\i}}n},
  {Guenther}, {Kenworthy}, \& {Zapatero Osorio}}]{pavlenko07lp944}
{Pavlenko}, Y.~V., {Jones}, H.~R.~A., {Mart{\'{\i}}n}, E.~L., {et~al.} 2007,
  MNRAS, 380, 1285

\bibitem[{{Pavlenko} \& {Schmidt}(2015)}]{pavlenko15}
{Pavlenko}, Y.~V., \& {Schmidt}, M. 2015, Kinematics and Physics of Celestial
  Bodies, 31, 90

\bibitem[{{Phan-Bao} {et~al.}(2009){Phan-Bao}, {Lim}, {Donati}, {Johns-Krull},
  \& {Mart{\'{\i}}n}}]{phanbao09}
{Phan-Bao}, N., {Lim}, J., {Donati}, J.-F., {Johns-Krull}, C.~M., \&
  {Mart{\'{\i}}n}, E.~L. 2009, ApJ, 704, 1721

\bibitem[{{Piskunov} {et~al.}(1995){Piskunov}, {Kupka}, {Ryabchikova}, {Weiss},
  \& {Jeffery}}]{piskunov95vald}
{Piskunov}, N.~E., {Kupka}, F., {Ryabchikova}, T.~A., {Weiss}, W.~W., \&
  {Jeffery}, C.~S. 1995, 112, 525

\bibitem[{{Quirrenbach} {et~al.}(2010){Quirrenbach}, {Amado}, {Mandel},
  {Caballero}, {Mundt}, {Ribas}, {Reiners}, {Abril}, {Aceituno}, {Afonso},
  {Barrado y Navascues}, {Bean}, {B{\'e}jar}, {Becerril}, {B{\"o}hm},
  {C{\'a}rdenas}, {Claret}, {Colom{\'e}}, {Costillo}, {Dreizler},
  {Fern{\'a}ndez}, {Francisco}, {Galad{\'{\i}}}, {Garrido}, {Gonz{\'a}lez
  Hern{\'a}ndez}, {Gu{\`a}rdia}, {Guenther}, {Guti{\'e}rrez-Soto}, {Joergens},
  {Hatzes}, {Helmling}, {Henning}, {Herrero}, {K{\"u}rster}, {Laun}, {Lenzen},
  {Mall}, {Martin}, {Mart{\'{\i}}n-Ruiz}, {Mirabet}, {Montes}, {Morales},
  {Morales Mu{\~n}oz}, {Moya}, {Naranjo}, {Rabaza}, {Ram{\'o}n}, {Rebolo},
  {Reffert}, {Rodler}, {Rodr{\'{\i}}guez}, {Rodr{\'{\i}}guez Trinidad},
  {Rohloff}, {S{\'a}nchez Carrasco}, {Schmidt}, {Seifert}, {Setiawan},
  {Solano}, {Stahl}, {Storz}, {Su{\'a}rez}, {Thiele}, {Wagner}, {Wiedemann},
  {Zapatero Osorio}, {del Burgo}, {S{\'a}nchez-Blanco}, \&
  {Xu}}]{quirrenbach10}
{Quirrenbach}, A., {Amado}, P.~J., {Mandel}, H., {et~al.} 2010, in SPIE, Vol.
  7735, SPIE, 13

\bibitem[{{Reid} \& {Hawley}(2005)}]{reid05}
{Reid}, I.~N., \& {Hawley}, S.~L. 2005, {New light on dark stars : red dwarfs,
  low-mass stars, brown dwarfs}, doi:10.1007/3-540-27610-6

\bibitem[{{Reiners} \& {Basri}(2007)}]{reiners07magnetic}
{Reiners}, A., \& {Basri}, G. 2007, ApJ, 656, 1121

\bibitem[{{Reiners} \& {Basri}(2009)}]{reiners09topology}
---. 2009, A\&A, 496, 787

\bibitem[{{Reiners} \& {Basri}(2010)}]{reiners10activity}
---. 2010, ApJ, 710, 924

\bibitem[{{Ribas}(2003)}]{ribas03lp944}
{Ribas}, I. 2003, A\&A, 400, 297

\bibitem[{Rice {et~al.}(1989)Rice, Wehlau, \& Khokhlova}]{rice89}
Rice, J.~B., Wehlau, W.~H., \& Khokhlova, V.~L. 1989, A\&A, 208, 179

\bibitem[{{Rockenfeller} {et~al.}(2006){Rockenfeller}, {Bailer-Jones}, \&
  {Mundt}}]{rockenfeller06mdwarfs}
{Rockenfeller}, B., {Bailer-Jones}, C.~A.~L., \& {Mundt}, R. 2006, A\&A, 448,
  1111

\bibitem[{{Ryabchikova} {et~al.}(2011){Ryabchikova}, {Pakhomov}, \&
  {Piskunov}}]{ryabchikova11vald}
{Ryabchikova}, T.~A., {Pakhomov}, Y.~V., \& {Piskunov}, N.~E. 2011, Kazan
  Izdatel Kazanskogo Universiteta, 153, 61

\bibitem[{{Stassun} {et~al.}(2012){Stassun}, {Kratter}, {Scholz}, \&
  {Dupuy}}]{stassun12radii}
{Stassun}, K.~G., {Kratter}, K.~M., {Scholz}, A., \& {Dupuy}, T.~J. 2012, ApJ,
  756, 47

\bibitem[{{Thibault} {et~al.}(2012){Thibault}, {Rabou}, {Donati},
  {Desaulniers}, {Dallaire}, {Artigau}, {Pepe}, {Micheau}, {Vall{\'e}e},
  {Pepe}, {Barrick}, {Reshetov}, {Hernandez}, {Saddlemyer}, {Pazder},
  {Par{\`e}s}, {Doyon}, {Delfosse}, {Kouach}, \& {Loop}}]{thibault12spirou}
{Thibault}, S., {Rabou}, P., {Donati}, J.-F., {et~al.} 2012, in SPIE, Vol.
  8446, SPIE, 30

\bibitem[{{Tuomi} {et~al.}(2014){Tuomi}, {Jones}, {Barnes},
  {Anglada-Escud{\'e}}, \& {Jenkins}}]{tuomi14mdwarfs}
{Tuomi}, M., {Jones}, H.~R.~A., {Barnes}, J.~R., {Anglada-Escud{\'e}}, G., \&
  {Jenkins}, J.~S. 2014, MNRAS, 441, 1545

\bibitem[{Unruh \& Collier~Cameron(1995)}]{unruh95profile}
Unruh, Y.~C., \& Collier~Cameron, A. 1995, MNRAS, 273, 1

\bibitem[{{Williams} {et~al.}(2015){Williams}, {Berger}, {Irwin},
  {Berta-Thompson}, \& {Charbonneau}}]{williams15}
{Williams}, P.~K.~G., {Berger}, E., {Irwin}, J., {Berta-Thompson}, Z.~K., \&
  {Charbonneau}, D. 2015, ApJ, 799, 192

\bibitem[{{Zerbi} {et~al.}(2014){Zerbi}, {Bouchy}, {Fynbo}, {Maiolino},
  {Piskunov}, {Rebolo Lopez}, {Santos}, {Strassmeier}, {Udry}, {Vanzi}, {Riva},
  {Basden}, {Boisse}, {Bonfils}, {Buscher}, {Cabral}, {Dimarcantonio}, {Di
  Varano}, {Henry}, {Monteiro}, {Morris}, {Murray}, {Oliva}, {Parry}, {Pepe},
  {Quirrenbach}, {Rasilla}, {Rees}, {Stempels}, {Valenziano}, {Wells}, {Wildi},
  {Origlia}, {Allende Prieto}, {Chiavassa}, {Cristiani}, {Figueira},
  {Gustafsson}, {Hatzes}, {Haehnelt}, {Heng}, {Israelian}, {Kochukhov},
  {Lovis}, {Marconi}, {Martins}, {Noterdaeme}, {Petitjean}, {Puzia}, {Queloz},
  {Reiners}, \& {Zoccali}}]{zerbi14hires}
{Zerbi}, F.~M., {Bouchy}, F., {Fynbo}, J., {et~al.} 2014, in Society of
  Photo-Optical Instrumentation Engineers (SPIE) Conference Series, Vol. 9147,
  Society of Photo-Optical Instrumentation Engineers (SPIE) Conference Series,
  23

\end{thebibliography}

\end{document}